\documentclass[12pt]{article}
\usepackage{fullpage}
\usepackage{setspace}
\usepackage{comment}
\usepackage[pdftex]{graphicx}
\usepackage{rotating}
\usepackage{amsmath}
\usepackage{amsthm}
\usepackage{amssymb}
\usepackage{bbm}
\usepackage{graphicx}
\usepackage{fullpage}
\usepackage{setspace}
\usepackage[numbers, authoryear]{natbib}
\usepackage{color}
\usepackage{soul}
\usepackage{booktabs}
\usepackage[linesnumbered, ruled]{algorithm2e}
\SetKwRepeat{Do}{do}{while}%

\usepackage{tikz, animate}
\usepackage{pgf}
\usepackage{pgfplots}
\usepackage{textpos}
\usepackage{graphicx}
\usepackage{pstricks}
\usepackage{framed}
\usepackage{varwidth}
\usepackage[normalem]{ulem}
\usepackage{xstring}
\usetikzlibrary{shapes}
\usetikzlibrary{arrows,decorations.pathmorphing,backgrounds,fit,positioning,shapes.symbols,chains}

\newcommand{\T}{\intercal}

\newcommand{\pn}{\mathbb{P}_{n}}

\newcommand{\OMIT}[1]{\relax}   
\def\text{{\rm}}

\def\dim{{\text{dim}}}

 \newcommand{\bma}[1]{\mbox{\boldmath $#1$}}

 \newcommand{\bX}{ {\bma{X}} }
 \newcommand{\bx}{ {\bma{x}} }

\newtheorem{thm}{Theorem}[section]
\newtheorem{lem}[thm]{Lemma}

\theoremstyle{definition}

\begin{document}

\title{Functional feature construction for individualized treatment
regimes}
\author{Eric B. Laber\\ Department of Statistics\\ North Carolina State
University
  \and Ana-Maria Staicu \\ Department of Statistics\\ North Carolina
State University}


\pagestyle{empty}
\begin{center}
  \textbf{Generalization error for decision problems (stat07998)} \\
  \textbf{Eric B. Laber$^1$ and Min Qian$^2$}
  \\
$^1$Department of Statistics, North Carolina State University,
  Raleigh, NC, 27695, U.S.A.  \\
$^2$Department of Biostatistics, Columbia University, 
New Yok NY 10032, U.S.A.
\end{center}

\begin{abstract}\noindent
In this entry we review the generalization error for classification
and single-stage decision problems.  
We distinguish three alternative definitions
of the generalization error which have, at times, been conflated in the
statistics literature and show that these definitions need not be
equivalent even asymptotically.   Because the generalization error is 
a non-smooth functional of the underlying generative model, standard
asymptotic approximations, e.g., the bootstrap or normal
approximations,  cannot guarantee correct frequentist
operating characterists without modification.  We provide
simple data-adaptive procedures that can be used to construct
asymptotically valid confidence sets for the generalization error.
We conclude the entry with a discussion of extensions and 
related problems.  
\end{abstract}

\setcounter{page}{1}
\pagestyle{plain}
\section{Introduction}
The generalization error of a predictive model is a measure of its
predictive performance when applied to make predictions in a
population of interest. In a wide range of supervised and
reinforcement learning problems, the estimated generalization error is
the primary means of comparing competing methodologies in a given
application domain or to benchmark competing algorithms across a suite of
test problems.  In prediction problems where the goal is to map inputs
into a finite set of predictive values, e.g., predicting a label in
classification or predicting an optimal treatment in precision
medicine, commonly used measures of generalization error can be
expressed as a weighted expected number of mistakes in the population
of interest, e.g., the (weighted) misclassification rate in
classification or the marginal mean outcome in precision medicine.  In
such settings, the estimated generalization error can be highly
sensitive to small perturbations in the data thereby making it difficult to
approximate its sampling distribution using standard asymptotic
methods \citep[][]{van1991differentiable, zhang1995ape,
  schiavo2000ten, laber2011adaptive, hirano2012,
  laber2014statistical}. 

Further complicating statistical inference is that the term
 `generaliation error' for a given
performance metric, e.g., the misclassification rate, may refer
to one of several funcationals of the underlying generative model,
the fitted data, and the algorithm used for estimation.  For example,
one measure of generalization error is the performance of the optimal 
predictive model within a pre-specified class; a  second is the 
conditional expected performance of a fitted model given the observed
data; and yet a third is the unconditional expected peformance of a fitted
model averaging over the observed data.  In this entry, we delineate
these three types of generalization error in the context of classifciation
and single-stage decision making.  We derive confidence intervals for the
generlization error in classification and then show that these methods
can be directly ported to the generalization error for decision 
problems.  Some the proposed methods are new  
including an adaptive projection interval designed to reduce
conservatism of standard projection intervals and a bounding method
that provides asymptotically correct conditional coverage given
the estimated optimal predictive model.  

An outline of the remainder of this entry is as follows.  In Section
2, we define three types of generalization error in classification and
show that they need not be close in arbitrarily large (but finite)
samples even under regularity conditions that make the estimators of
these quantities asymptotically normal and pointwise convergent to the
same limit.  We then derive inference procedures under a moving
parameter asymptotic framework that accounts for this finite sample
behavior. In Section 3, we exploit the fact that the marginal mean
outcome in a single-stage decision problem can be recast as a weighted
misclassification error and show how the methods developed for
classification can thereby be directly extended.  We provide
concluding remarks in Section 4.


\section{Generalization error in classification}
\label{sec:classification}
We assume that the observed data are $\left\lbrace 
\left(\bX_i, Y_i\right)\right\rbrace_{i=1}^{n}$ 
which comprise $n$ independent replicates of the
input--label pair $(\bX, Y)$, 
 where the input, $\bX$, 
takes values in $\mathbb{R}^p$ and the label, $Y$,  is binary
and coded to take values in $\left\lbrace -1,1\right\rbrace$.  
A classification rule is map, $c:\mathrm{dom}\,\bX \rightarrow
\mathrm{dom}\,Y$ so that under $c$ the predicted label at
input $\bX=\bx$ is $c(\bx)$. Let $P$ denote joint
distribution of $(\bX, Y)$, then the generalization error 
of a rule $c$ is 
\begin{equation*}
M(c) \triangleq P\mathbbm{1}\{Y\ne c(\bX)\} = 
P\mathbbm{1}\{Yc(\bX) < 0\},
\end{equation*}
where $\mathbbm{1}\{u\}$ is an indicator that the condition $u$ is
true. Thus, $M(c)$ captures the probability $c$ will incorrectly
label randomly the input $\bX$ of the 
input-label pair $(\bX, Y)$ randomly drawn from $P$.  
Let $\pn$ denote the empirical distribution, then the plug-in estimator
of $M(c)$ is $\widehat{M}_{n}(c) \triangleq \pn \mathbbm{1}\left\lbrace
Yc(\bX) < 0
\right\rbrace$.  For a fixed classification rule, $c$, it follows
from the central limit theorem that
$\sqrt{n}\left\lbrace \widehat{M}_{n}(c) - M(c)\right\rbrace
\leadsto \mathrm{Normal}\left[
0, M(c)\left\lbrace  1- M(c)\right\rbrace
\right]$.  Hence, estimation of and inference for the generalization error
of a fixed classification rule is straightforward.  However, when
estimation and inference are focused on optimal classification in
a given problem domain, quantifying uncertainty about the generalization
error is considerably more complex.  


Let $\Omega$ denote the space of distributions over
$\mathrm{dom}\left(\bX\times Y\right)$ and define a classification
algorithm for the class $\mathcal{C}$ to be a map
$\Gamma: \Omega\rightarrow \mathcal{C}$.  Under $\Gamma$, the
estimated optimal classifier given the observed data
$\left\lbrace (\bX_i, Y_i)\right\rbrace_{i=1}^{n}$ is
$\widehat{c}_{n} = \Gamma(\pn)$.  The generalization error of
 $\widehat{c}_{n}$ is $M(\widehat{c}_{n})$, i.e., the missclassication rate
associated with applying $\widehat{c}_{n}$ to a new input-label pair
$(\bX, Y)$ drawn from $P$.  Thus, the generalization error is useful
for quantifying the value of applying the estimated classification
rule, $\widehat{c}_{n}$, to make decisions in the domain of interest.
Because
$\widehat{c}_{n}$ is a function of both the observed data and the
unknown distribution $P$, it is an example of a data-dependent
parameter \citep[][]{dawid1994selection, efron1997improvements, 
laber2011adaptive}.  While data-dependent parameters are somewhat
unusual, the definition of a confidence set for a such a parameter 
closely matches that of a fixed (i.e., not data-dependent) parameter.
Given $\alpha \in (0,1)$ we say that the set 
$\widehat{S}_{n,1-\alpha} = S_{n,1-\alpha}(\pn)$ 
is a
$(1-\alpha)\times 100\%$ asymptotic 
confidence set for $M(\widehat{c}_{n})$ if
$P\left(
M(\widehat{c}_{n}) \in \widehat S_{n, 1-\alpha}
\right)  \ge 1-\alpha + o(1)$, where the probability statement
is over both a new draw $(\bX, Y)$ and the distribution of
the observed data. Alternatively, one can define 
a $(1-\alpha)\times 100\%$ 
 asymptotic conditional confidence set, say $\widehat{H}_{n,1-\alpha} = 
H_{n,1-\alpha}(\pn)$, which ensures that
$P\left\lbrace M(\widehat{c}_{n}) \in \widehat{H}_{n,1-\alpha}
\big|\,\widehat{c}_{n}\right\rbrace \ge 1-\alpha + o_P(1)$. It
can be seen that a conditional confidence set is also a non-conditional
confidence which is an appealing feature; however, the construction
of such sets is not always straightforward 
\citep[][]{casella1992conditional, robins2014discussion}.

Given a classification algorithm $\Gamma$,
define $c^{\mathrm{opt}} = \Gamma(P)$ to be
the optimal classification rule relative to $\Gamma$. Thus,
$c^{\mathrm{opt}}$ need not be the optimal classifier over the space
of all measurable maps from $\mathrm{dom}\,\bX$ into
$\mathrm{dom}\,Y$, i.e., the Bayes classifier which is given by
$c^{\mathrm{Bayes}}(\bx) = \mathrm{sign}\left\lbrace P(Y=1|\bX=\bx) -
  1/2 \right\rbrace$
\citep[][]{duda2012pattern}.  Indeed, $c^{\mathrm{opt}}$ need not
even be the optimal regime in the class of regimes defined
by the image of $\Gamma$.  
Nevertheless,
we define the population-optimal
generalization error to be
$M(c^{\mathrm{opt}})$.  Inference for $M(c^{\mathrm{opt}})$ may be of
interest in the context of evaluating the benefit of a data-driven
classification rule relative to some existing {\em ad hoc} classifier
that is already in place.  For example, consider the problem of
predicting treatment response among of breast cancer patients; one may
wish to know if a data-driven classification rule based on
high-dimensional gene expression signatures outperforms an existing
classifier based on a much smaller subset of biomarkers
\citep[][]{luckett2018ROC}. In this example, if $c_0$ is the existing
classification rule, one might be interested in testing
$H_0: M(c_0) \le M(c^{\mathrm{opt}})$.  Thus, the population-optimal
generalization error is useful for quantifying the value of a 
classification rule in a given domain.  However, this does not
measure the quality of a classification rule estimated from a finite
data-set which is of greater interest if a data-driven rule is to 
be deployed in a given application domain.

An alternative measure of performance is the expected generalization
error of a learning algorithm defined as
$M_{n}(\Gamma) = \mathbb{E}M(\widehat{c}_{n})$ which is the average
performance of the classification algorithm $\Gamma$ across $i.i.d.$
samples of size $n$ drawn from $P$.  The curve,
$n\mapsto M_{n}(\Gamma)$, known as the learning curve of the algorithm
$\Gamma$, is a measure of how efficiently the algorithm learns from
data on average \citep[][]{amari1992four, haussler1996rigorous,
  mukherjee2003estimating, laber2016imputation}; however, for the 
purpose of sample size calculation intended to ensure high-quality estimation
of an optimal classification rule, percentiles of the sampling
distribution of $M(\widehat{c}_{n})$ as a function of $n$
may be more scientifically meaningful \citep[][]{laber2016using}. 
The quantity $M_n(\Gamma)$ (or the learning curve) is most useful as a 
means to compare algorithms in a given domain across a range of 
data set sizes.

The three generalization errors $M(c^{\mathrm{opt}})$, 
$M(\widehat{c}_n)$, and $M_{n}(\Gamma)$ represent three different
performance metrics. In the next section, we characterize the
asymptotic behavior of these quantities in the special case of a linear classifier fit using least squares. Using this simple example, we show, 
perhaps somewhat
surprisingly, that the three generalization errors need
not converge to each other even as $n$ diverges to $\infty$.

\subsection{Asymptotic behavior of the generalization error for
linear classification rules} 
For the purpose of illustrating the asymptotic behavior of the three
types of generalization error defined in the preceding section, we
consider linear classifiers fit using least squares; these results
extend with minor modification to other convex loss functions, e.g.,
logistic, hinge, exponential among others \citep[see,
e.g.,][]{laber2011adaptive}.  Define
$\widehat{\beta}_{n} = \arg\min_{\beta\in\mathbb{R}^p} \pn\left(
  Y-\bX^{\T}\beta\right)^2$
and
$\widehat{c}_{n}(\bx) = \mathrm{sign}(\bx^{\T}\widehat{\beta}_{n})$.
Let $\beta^* = \arg\min_{\beta\in\mathbb{R}^p}
P\left(Y-\bX^{\T}\beta\right)^2$ denote the population analog of
$\widehat{\beta}_{n}$ and define $c^{\mathrm{opt}}(\bx) = 
\mathrm{sign}\left(\bx^{\T}\beta^*\right)$.\footnote{Note that
$c^{\mathrm{opt}}(\bx)$ need not equal 
$\arg\min_{c\in\mathcal{C}_{\mathrm{Lin}}}M(c)$, where
$\mathcal{C}_{\mathrm{Lin}}$ is the space of linear classifiers; this
is because of  
the mismatch between least squares and zero-one loss.  
See \citep[][]{bartlett2006convexity, mintron}.}.  
Under this model, the generalizations are:
$M(c^{\mathrm{opt}}) = P\mathbbm{1}\left\lbrace
Y\bX^{\T}\beta^* < 0
\right\rbrace$; $M(\widehat{c}_{n}) \allowbreak =\allowbreak  P
\mathbbm{1}\left\lbrace Y\bX^{\T}\widehat{\beta}_{n} < 0
\right\rbrace \allowbreak = \\  \allowbreak \int 
\mathbbm{1}\left\lbrace y\bx^{\T}\widehat{\beta}_{n} < 0
\right\rbrace dP(\bx, y)$; and
$M_{n}(\Gamma) = \mathbb{E}M(\widehat{c}_{n})$. Under
 mild moment conditions, $\sqrt{n}(\widehat{\beta}_{n}
-\beta^*) \leadsto \mathrm{Normal}\left\lbrace 
0, \Sigma(\beta^*)\right\rbrace$, where
$\Sigma(\beta^*) = \left( P\bX\bX^{\T}\right)^{-1}
P(Y-\bX^{\T}\beta^*)^2\bX\bX^{\T}
\left( P\bX\bX^{\T}\right)^{-1}$
\citep[e.g.,][]{stefanski2002calculus, seber2012linear}.  
Thus, it follows that 
\begin{eqnarray*}
M(\widehat{c}_{n}) &=& P\mathbbm{1}\left\lbrace
Y\bX^{\T}\widehat{\beta}_n < 0
\right\rbrace \\
&=& P\mathbbm{1}\left\lbrace
Y\bX^{\T}\sqrt{n}\left(\widehat{\beta}_{n}-\beta^*\right) < 0
\right\rbrace \mathbbm{1}\left\lbrace
\bX^{\T}\beta^* = 0 
\right\rbrace \\ 
&&\quad + \quad  P\mathbbm{1}\left\lbrace
Y\bX^{\T}\beta^* < 0
\right\rbrace\mathbbm{1}\left\lbrace \bX^{\T}\beta^* \ne 0 \right\rbrace
 + o_P(1) \\
&\leadsto & P\mathbbm{1}\left\lbrace
Y\bX^{\T}\mathbb{Z} < 0
\right\rbrace \mathbbm{1}\left\lbrace
\bX^{\T}\beta^* = 0 
\right\rbrace  + P\mathbbm{1}\left\lbrace
Y\bX^{\T}\beta^*< 0
\right\rbrace,
\end{eqnarray*} 
where
$\mathbb{Z} \sim \mathrm{Normal}\left\lbrace 0, \Sigma(\beta^*)
\right\rbrace$
and we have used
$P\mathbbm{1}\left\lbrace Y\bX^{\T}\beta^* < 0
\right\rbrace\mathbbm{1}\left\lbrace \bX^{\T}\beta^* \ne 0
\right\rbrace = P\mathbbm{1}\left\lbrace Y\bX^{\T}\beta^*< 0
\right\rbrace$.
Using the above expression and applying the dominated convergence to
interchange limits and expectations, it follows that
$M_{n}(\Gamma) = \mathbb{E}M(\widehat{c}_{n}) \rightarrow
P\mathbbm{1}\left\lbrace \bX^{\T}\beta^* = 0\right\rbrace/2 +
P\mathbbm{1}\left\lbrace Y\bX^{\T}\beta^*< 0 \right\rbrace$
as $n\rightarrow \infty$.  Thus, we see that the three types of
generalization error, $M(c^{\mathrm{opt}})$, $M(\widehat{c}_{n})$, and
$M_{n}(\Gamma)$, need not coincide even asymptotically.

In the preceding derivations, the differences across the three types
of generalization error depend on the amassing of data points on the
boundary $\bx^{\T}\beta^* = 0$; indeed, if $P(\bX^{\T}\beta^* = 0) = 0$
then the three definitions converge to the same limit. Thus, it may be
tempting in conducting inference to assume---perhaps even correctly,
e.g., if $\bX$ is continuous and $\beta^*$ is not identically
zero---that $P(\bX^{\T}\beta^* = 0) = 0$.  Unfortunately, the abrupt
dependence on this probability 
in the limiting behavior of the generalization error
 is a symptom of nonregular behavior which manifests in a lack
of uniform convergence. The practical consequences of this lack of
uniformity include that: (i) the three measures of generalization
error need not be close in finite samples; and (ii) standard
asymptotic approximations to the sampling distributions of estimators
of the generalization, e.g., normal approximations, can perform poorly
in finite samples.  To illustrate the first point, consider
the following generative model: $Y\sim \mathrm{Uniform}\lbrace -1,
1\rbrace$, $U\sim\mathrm{Uniform}[0,1]$, 
$X|Y=1\, \sim \mathrm{Normal}
\left(2 - 4*\mathbbm{1}\left\lbrace U\le 1/2-\delta\right\rbrace, 
\sigma^2 \right)$, and $X|Y=-1\,\sim \mathrm{Normal}(0,0.5^2)$, which
indexed by the parameter $\delta\in [0,1/2]$. Thus, 
if $Y=1$ then the input, $X$, follows a mixture of two normal 
distributions with component means $-2$ and $2$, 
component variances both equal  to
$0.5^2$, and 
mixture probabilities $1/2-\delta$ and $1/2 + \delta$.  We 
consider linear decision rules of the form 
$c(x) = \mathrm{sign}\left(\beta_0 + \beta_1 x\right)$ indexed
by $\beta = (\beta_0, \beta_1)^{\T}$.  It can be seen that if 
$\delta = 0$ then $\beta^* = \arg\min_{\beta}P(Y-\beta_0 -\beta_1 X)^2$
is identically zero; otherwise, $\beta^*$ is nonzero and because
$X$ is continuous $P(\beta_0^* + \beta_1^*X = 0) = 0$.  
We draw a sample of size 125 from the distribution 
of $M(\widehat{c}_{n})$ under this model for training set 
sizes ranging from $n=50$ to $n=50000$.  This generative model illustrates that  $M(\widehat{c}_n)$ can be unstable even in large samples; this 
lack of stability is due to the presence of the non-smooth indicator
function in the definition of $M(\widehat{c}_{n})$ 
which is sensitive to small perturbations of its
arguments.  

\begin{figure}[p]
\begin{minipage}{0.45\linewidth}
\includegraphics[width=1.0\linewidth]{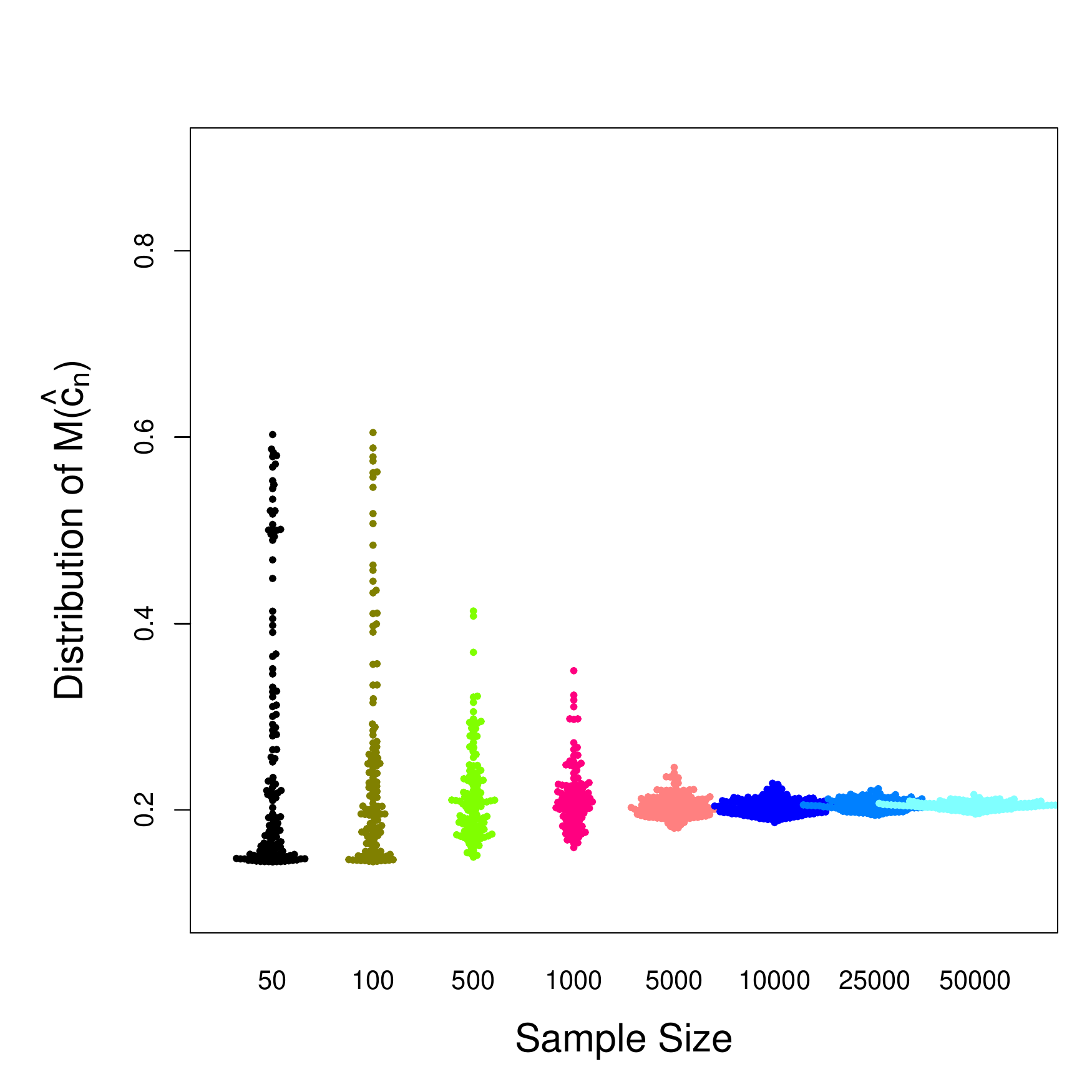}
\end{minipage}
\hspace{0.05\linewidth}
\begin{minipage}{0.45\linewidth}
\includegraphics[width=1.0\linewidth]{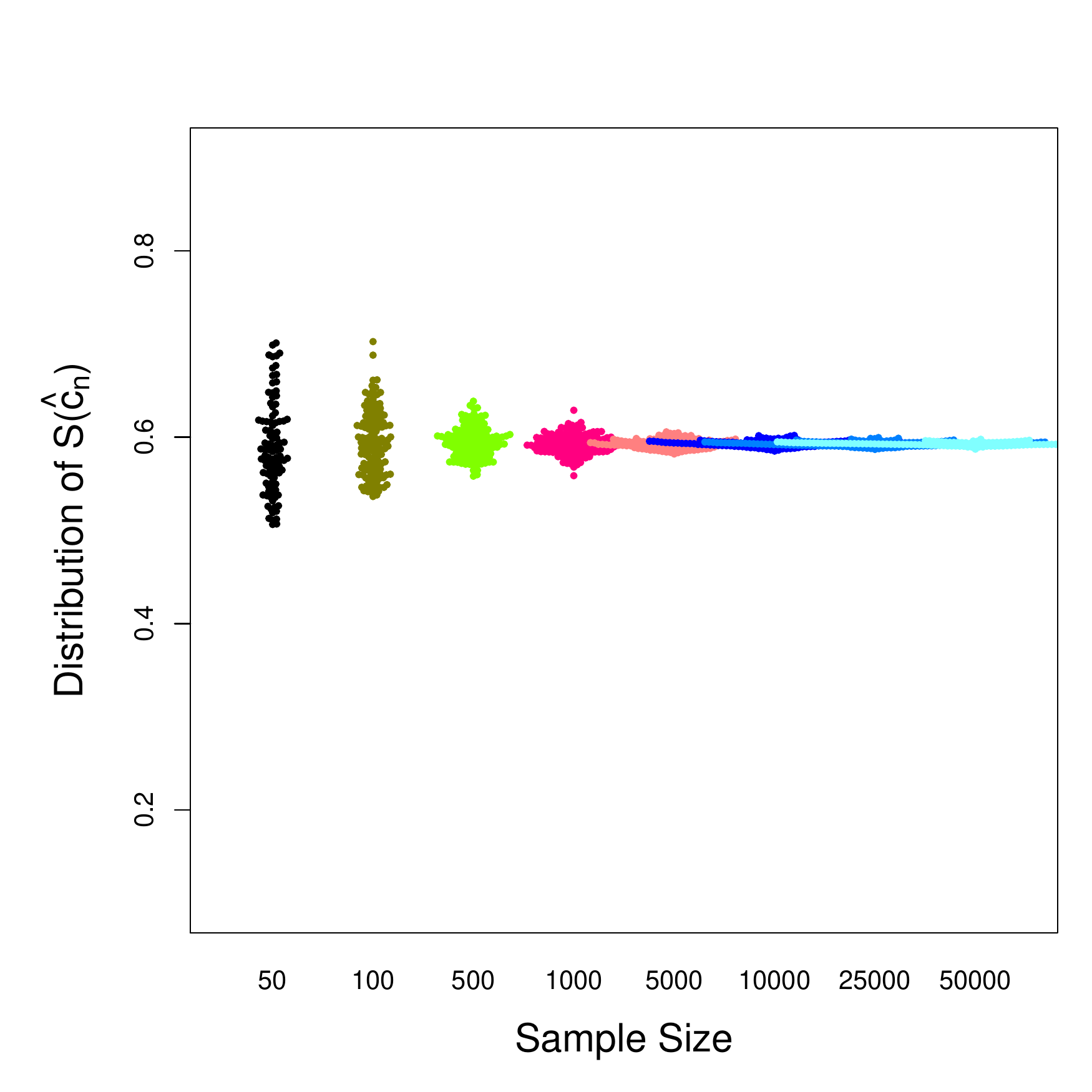}
\end{minipage}\\
\begin{minipage}{0.45\linewidth}
\includegraphics[width=1.0\linewidth]{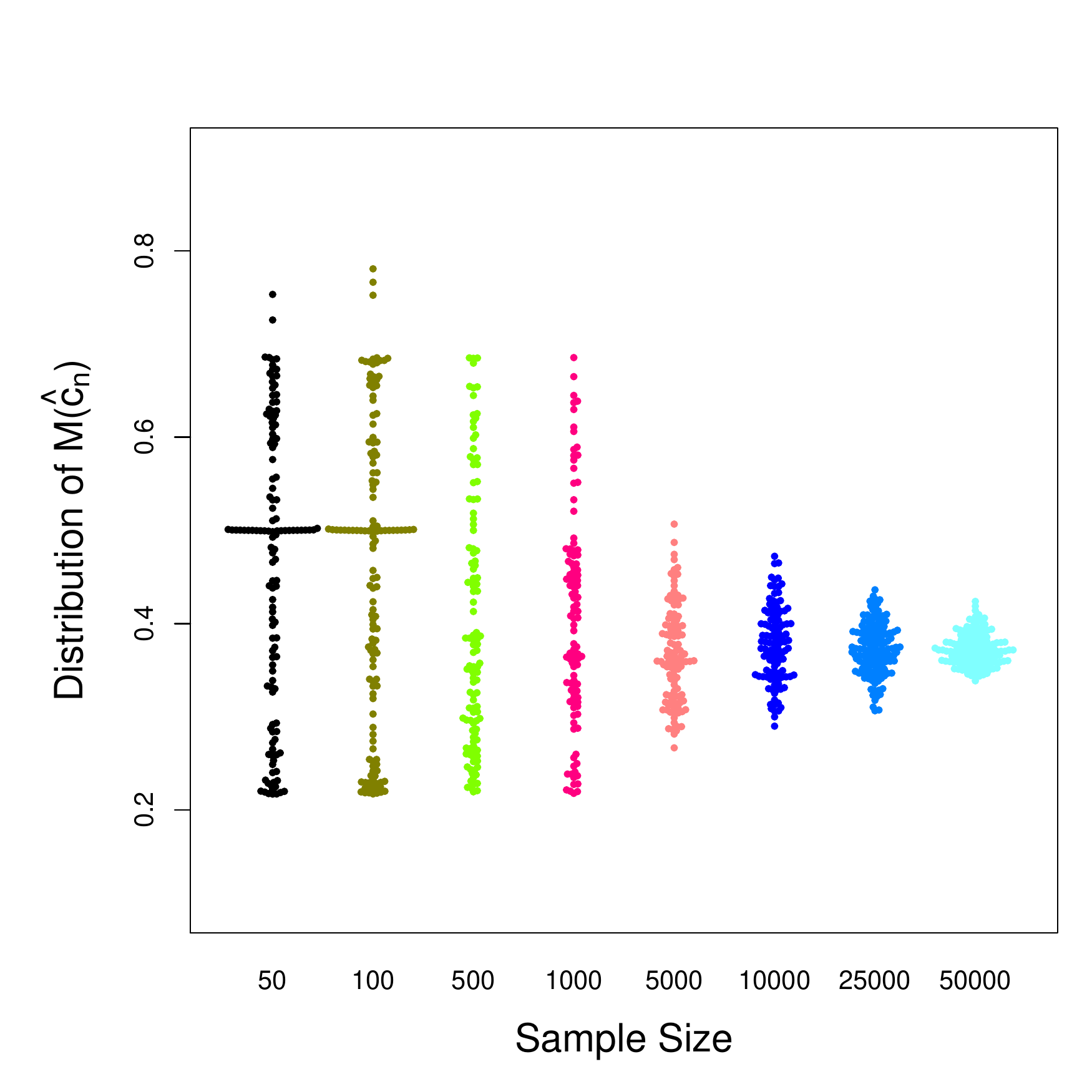}
\end{minipage}
\hspace{0.05\linewidth}
\begin{minipage}{0.45\linewidth}
\includegraphics[width=1.0\linewidth]{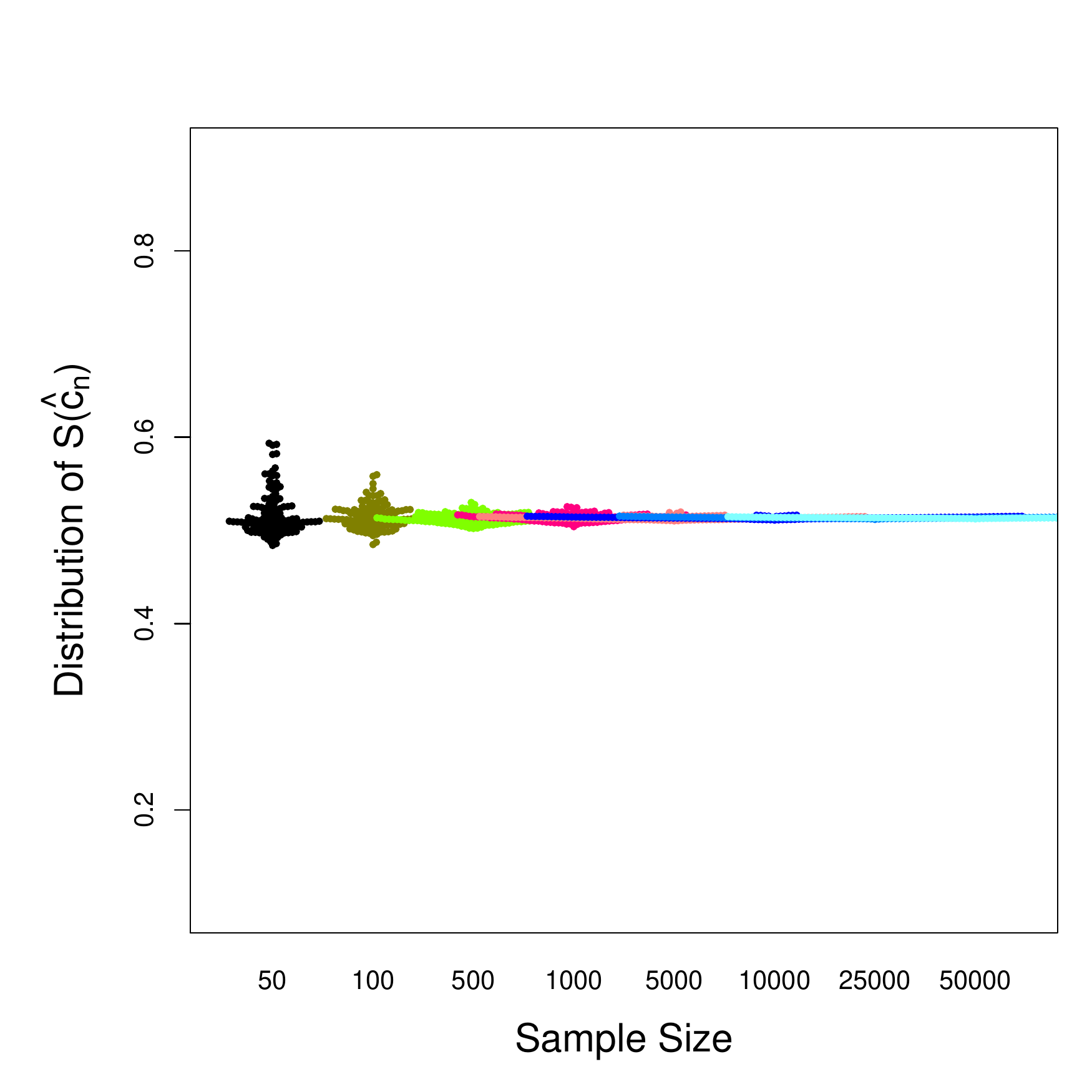}
\end{minipage}\\
\begin{minipage}{0.45\linewidth}
\includegraphics[width=1.0\linewidth]{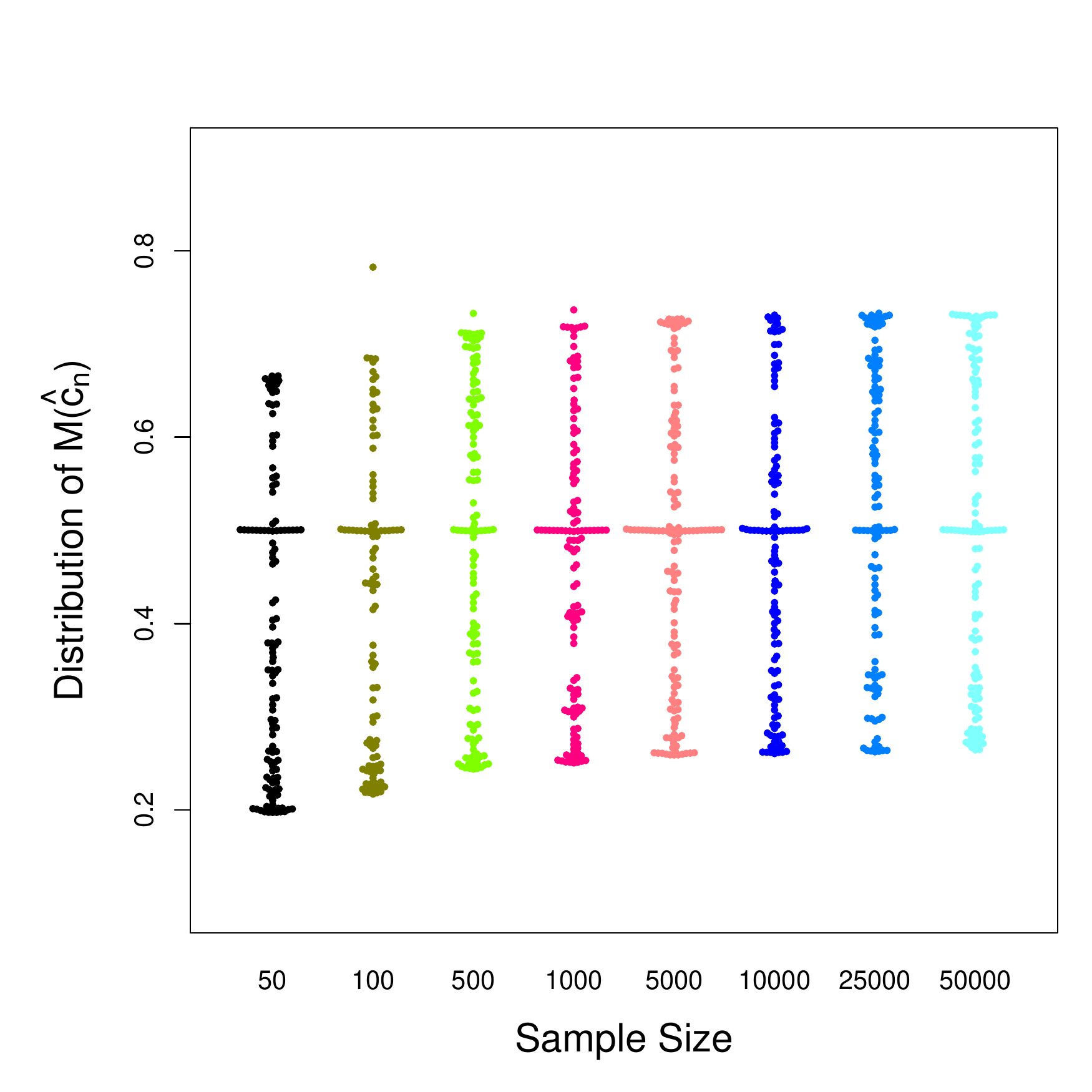}
\end{minipage}
\hspace{0.05\linewidth}
\begin{minipage}{0.45\linewidth}
\includegraphics[width=1.0\linewidth]{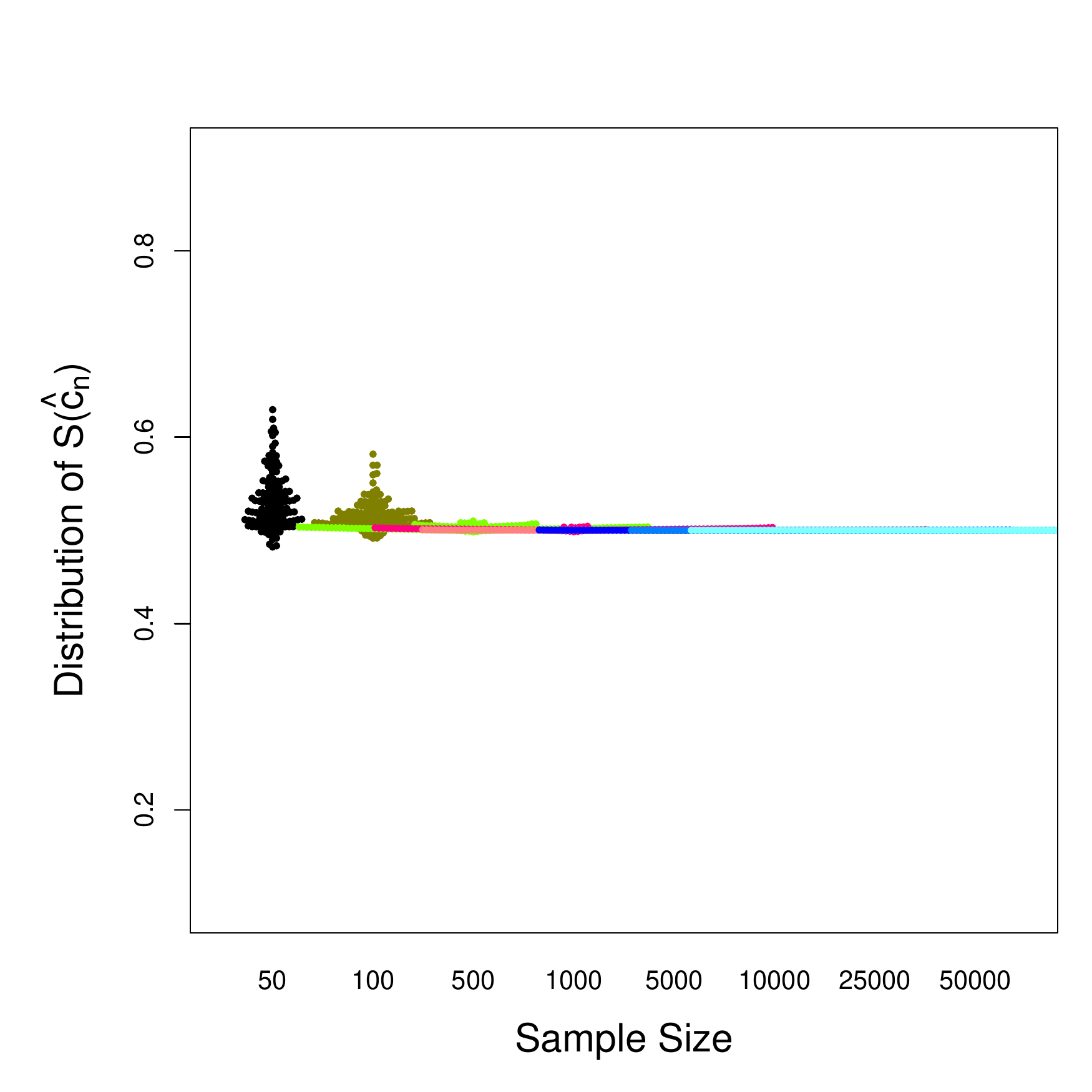}
\end{minipage}

\caption{
\label{genErrDistn} 
\textbf{Left column (from top to bottom):} 1D histogram of 
$M(\widehat{c}_{n})$ by training set size for $\delta=0.25$,
$\delta=0.10$, and $\delta=1/\sqrt{n})$.  
\textbf{Right
column (from top to bottom):} 1D histogram of $S_{\tau}(\widehat{c}_{n})$
for $\delta=0.25$, $\delta=0.10$, and $\delta=1/\sqrt{n}$.  
}
\end{figure}

To highlight the impact of this nonsmoothness,
we also draw a sample of size 125 drawn from the 
distribution of $S_{\tau}(\widehat{c}_{n}) \triangleq
P\mathrm{expit}\left\lbrace 
-\tau Y(\widehat{\beta}_{0,n} + 
\widehat{\beta}_{1,n} X)\right\rbrace$ where 
$\mathrm{expit}(u) = \exp(u)/\lbrace 1+\exp(u)\rbrace$ 
and $\tau > 0$.  
As $\tau\rightarrow \infty$ the function
$\mathrm{expit}(-\tau u)$ converges to 
$1_{u < 0}$; thus,
$S(\widehat{c}_{n})$ can 
be viewed as a smooth surrogate for $M(\widehat{c}_{n})$.   In
our simulated examples we set $\tau = 3$.   
When $\delta\ne 0$,
$M(\widehat{c}_{n})$ converges in probability to $M(c^{\mathrm{opt}})$,
where $c^{\mathrm{opt}}(\bx) = 
\mathrm{sign}\left(\beta_0^* + \beta_1^*\bx\right)$.  The top left
panel of Figure \ref{genErrDistn} shows a one-dimensional histogram
of the distribution of $M(\widehat{c}_{n})$ for 
$\delta = 0.25$ whereas the top right panel of this figure
shows one-dimensional histogram for the smooth surrogate 
 $S(\widehat{c}_{n})$ under the same generative model.
 The higher variability of 
 $M(\widehat{c}_{n})$ relative to 
 $S(\widehat{c}_{n})$ is striking as the $S(\widehat{c}_{n})$ 
 is as tightly clustered about its mean at a training set
 size of $n=500$ as $M(\widehat{c}_{n})$ is at $n=5000$. 
 This difference can be made more pronounced by choosing
 a smaller value of $\delta$.  The middle left and middle panels of 
 Figure \ref{genErrDistn} display one-dimensional histograms of
 $M(\widehat{c}_{n})$ and $S(\widehat{c}_{n})$ for $\delta =0.10$.
 In this case, $S(\widehat{c}_{n})$ is more tightly clustered
 about its mean at $n=100$ then $M(\widehat{c}_{n})$ is at 
 $n=50000$. Thus, using the asymptotic limit of 
 $M(\widehat{c}_{n})$ to approximate its finite
 sample behavior can be arbitrarily poor.  Indeed,
 the lower left and lower right panels of 
 Figure \ref{genErrDistn} show the one-dimensional histograms
 of $M(\widehat{c}_{n})$ and $S(\widehat{c}_{n})$ 
 when $\delta = 1/\sqrt{n}$; in this case, the variability
 of $M(\widehat{c}_{n})$ remains the same regardless of 
 how large the sample size grows whereas $S(\widehat{c}_n)$ quickly
 concentrates about its mean.  

 Letting the parameter $\delta$ change with the sample size is an
 example of a moving-parameter asymptotic analysis.  Moving parameter
 asymptotic analyses are commonly used to study the limiting behavior
 of nonregular quantities like $M(\widehat{c}_{n})$ as they retain
 salient small sample behaviors even in infinite samples.  In our
 example, we saw that letting $\delta = O(1/\sqrt{n})$ retained the
 small-sample instability of $M(\widehat{c}_{n})$ even as $n$
 diverged.  On the other hand, the concentration of 
$S(\widehat{c}_{n})$ about its mean was unaffected by letting 
the parameter $\delta$ vary with $n$.  The
 empirical behavior of $M(\widehat{c}_{n})$ and $S(\widehat{c}_{n})$
 is consistent with the formal definition of a nonregular parameter as
 being sensitive to local (i.e., $O(1/\sqrt{n})$) perturbations of the
 underlying generative model \citep[][]{tsiatis2007semiparametric}.
 Because we are interested in asymptotic methods that
faithfully reflect the finite sample behavior of the generalization
errors we shall consider moving asymptotic arguments in our
development of inferential methods. 

\subsubsection{Confidence intervals for $M(c^{\mathrm{opt}})$} 
For each $\beta\in\mathbb{R}^p$ write $M(\beta) = 
P\mathbbm{1}\left\lbrace Y\bX^{\T}\beta < 0\right\rbrace$ 
to denote the generalization error for the classification
rule $c(\bx) = \mathrm{sign}\left(\bx^{\T}\beta\right)$;
similarly, write $\widehat{M}_{n}(\beta) = \pn 
\mathbbm{1}\left\lbrace Y\bX^{\T}\beta < 0\right\rbrace$ to 
be the plugin estimator of $M(\beta)$.  
Define $\sigma^2(\beta) = M(\beta)\left\lbrace 1-M(\beta)\right\rbrace$
and $\widehat{\sigma}_{n}^2(\beta) = \widehat{M}_{n}(\beta)
\left\lbrace 1- \widehat{M}_{n}(\beta)\right\rbrace$.  
Were $\beta^*$ known, 
then, as noted previously,  
$\sqrt{n}\left\lbrace \widehat{M}_{n}(\beta^*) 
- M(\beta^*)\right\rbrace \leadsto 
\mathrm{Normal}\left\lbrace 0, 
\sigma^2(\beta)
\right\rbrace$. 
Thus, for any $\alpha \in (0,1)$ an asymptotic 
$(1-\alpha)\times 100\%$ confidence interval for $M(\beta^*)$ is 
\begin{equation*}
\mathfrak{Z}_{n,1-\alpha}(\beta^*) = 
\left[
\widehat{M}_{n}(\beta^*) -
\frac{
z_{1-\alpha/2}
\widehat{\sigma}_{n}(\beta^*)
}{
\sqrt{n}
},\, 
\widehat{M}_{n}(\beta^*) +
\frac{
z_{1-\alpha/2}
\widehat{\sigma}_{n}(\beta^*)
}{
\sqrt{n}
}
\right],
\end{equation*}
where $z_{\nu}$ is the $\nu\times 100$ percentile
of a standard normal distribution.  Of course $\beta^*$ is not 
generally known and must be estimated using the observed data.  
However, as we have seen, $\widehat{M}_{n}(\widehat{\beta}_{n})$
need not converge to $M(\beta^*)$, making
$\sqrt{n}\left\lbrace \widehat{M}_{n}(\widehat{\beta}_{n})
- M(\beta^*)\right\rbrace$ a tenuous starting point 
for inference.  Instead, we utilize the fact that
$\widehat{\beta}_{n}$ is regular and asymptotically normal 
to construct a projection confidence interval
\citep[][]{berger1994p, robinsTF}; because projection
intervals are notoriously conservative, we also present some
refinements intended to reduce conservatism.

 Define $\widehat{\Sigma}_{n}  \triangleq
 \left(\pn \bX\bX^{\T}\right)^{-1}\pn (Y-\bX^{\T}\widehat{\beta}_{n})^2
 \bX\bX^{\T}\left(\pn\bX\bX^{\T}\right)^{\T}$ to be an 
 estimator of the asymptotic variance-covariance matrix of 
 $\widehat{\beta}_{n}$.
 For any $\eta \in (0,1)$ a $(1-\eta)\times 100\%$ Wald-type 
 confidence set for $\beta^*$ is 
 $\mathfrak{F}_{n,1-\eta} \triangleq  
 \left\lbrace
 \beta\in\mathbb{R}^p\,:\, 
 n\left(\widehat{\beta}_{n} - \beta\right)^{\T}\widehat{\Sigma}_{n}^{-1}
 \left(
 \widehat{\beta}_{n} - \beta
 \right) \le \chi_{n, 1-\eta}^2
 \right\rbrace$, where
$\chi_{d, \nu}^2$ is the $\nu\times 100$ percentile of 
chi-square random variable with $d$ degrees of freedom.  
Let $\omega \in (0,1)$ and choose $\eta, \alpha\in(0,1)$ so
that $\omega = \alpha + \eta$. 
A $(1-\omega)\times 100\%$ projection interval for 
$M(c^{\mathrm{opt}}) = M(\beta^*)$ is 
$\mathfrak{P}_{n, 1-\omega} \triangleq 
\bigcup_{\beta \in \mathfrak{F}_{n, 1-\eta}}
\mathfrak{Z}_{n,1-\alpha}
(\beta)$.  To see that this provides the correct coverage
asymptotically write
\begin{eqnarray*}
P\left\lbrace 
M(\beta^*) \notin \mathfrak{P}_{n, 1-\omega}
\right\rbrace &=& 
P\left\lbrace 
M(\beta^*) \notin \mathfrak{P}_{n, 1-\omega},\,
\beta^*\in \mathfrak{F}_{n,1-\eta}
\right\rbrace +
P\left\lbrace 
M(\beta^*) \notin \mathfrak{P}_{n, 1-\omega},\,
\beta^*\notin \mathfrak{F}_{n,1-\eta}
\right\rbrace \\
&\le& P\left\lbrace 
M(\beta^*) \notin \mathfrak{Z}_{n,1-\alpha}(\beta*)\right\rbrace 
+ P\left( \beta^*\notin \mathfrak{F}_{n,1-\eta}\right)\\
&\le & \alpha + \eta + o(1)\\
&=& \omega + o(1). 
\end{eqnarray*}
Because the projection interval is constructed from
continuous operations with regular estimators it is insensitive
to local perturbations as we will show below.

To evaluate the local behavior of the generalization error we make
the following assumptions.
\begin{itemize}
  \item[(A1)] The distribution $P$ satisfies $P||Y||^2||\bX||^2 < \infty$.
  \item[(A2)] The covariance $\Sigma(\beta)$ is strictly positive definite
for all $\beta$ in a neighborhood of $\beta^*$.
  \item[(A3)] For any $\ell \in \mathbb{R}^p$ there exists a sequence
    of distributions $P_n$ that satisfies 
    \begin{equation*}
      \int \left[
        \sqrt{n}\left(
          dP_{n}^{1/2}-dP^{1/2}
          \right) - \frac{1}{2}v_{\ell}dP^{1/2}
        \right]^2 \rightarrow 0,
    \end{equation*}
    for some real-valued function $v_{\ell}$ for which:
    (i) if $\beta_n^* = \arg\min_{\beta}P_n\left(Y-\bX^{\T}\beta\right)^2$
      then $\beta_{n}^* = \beta^* + \ell/\sqrt{n} + o(1/\sqrt{n})$;
      and (ii) $P_n||Y||^2||\bX||^2$ is a uniformly bounded sequence. 
\end{itemize}
 
The following result states that probjection interval remains
valid even under a moving parameter asymptotic framework.
\begin{thm}
Assume (A1)-(A2) and that for each $n$ the observed
data are $\left\lbrace \left(\bX_{i,n}, Y_{i,n}\right)
\right\rbrace_{i=1}^{n}$ which comprise an $i.i.d.$
draw from $P_n$ which satisfies (A3).  
Let $M_n(\beta_n^*) = P_n\mathbbm{1}\left\lbrace
Y\bX^{\T}\beta_{n}^* < 0
\right\rbrace$,
then 
$P_n\left\lbrace M_n(\beta_{n}^*) \in \mathfrak{P}_{n,1-\omega}
\right\rbrace \ge 1-\omega + o(1).$ 
\end{thm}\noindent
A proof of this result follows from noting that 
$\sqrt{n}(\widehat{\beta}_{n} - \beta_{n}^*) \leadsto
\mathrm{Normal}\left\lbrace 0, \Sigma(\beta^*)\right\rbrace$
and $\sqrt{n}\left\lbrace \widehat{M}_{n}(\beta) -
M_n(\beta)\right\rbrace \leadsto \mathrm{Normal}\left\lbrace 
0,
\sigma^2(\beta)\right\rbrace$ for each fixed $\beta$
under $P_n$; thus, the result follows by applying the
argument given for the fixed parameter setting above.

Projection intervals are appealing because of their simplicity and
generality.  However, they can be extremely conservative in some
settings \citep[][]{laber2014statistical}.  One driver for this
conservatism is that the projection interval 
is not adaptive to the concentration of points
about the boundary $\bx^{\T}\beta^*=0$; e.g., if $\bX^{\T}\beta^*$
were bounded away from zero with probability one, then one could just
apply the standard nonparametric bootstrap or a normal-based
approximation to construct a confidence interval
\citep[][]{efron1994introduction, shao2012jackknife}.  One can reduce
conservatism by adapting the projection interval so that the union
only affects points that lie `near' to the decision boundary
$\bx^{\T}\beta^*$.  We formalize the notion of being near the boundary
using a hypothesis test of the null $H_0(\bx):\bx^{\T}\beta^* = 0$
against a two-sided alternative for each input $\bX= \bx$ in the observed
data.  Define the test statistic
$T_n(\bx) =
(\bx^{\T}\widehat{\beta}_n)^2/(\bx^{\T}\widehat{\Sigma}_n\bx)$;
we assume (A4) that $\bX^{\T}\widehat{\Sigma}_{n}\bX$ is bounded away
from zero with probability one.  We reject $H_0(\bx)$ if $T_n(\bx)$ is
large and fail to reject otherwise.  Let $\lambda_n$ denote a positive
sequence of critical values (see below for additional restrictions on this 
sequence) and define
\begin{equation*}
  \mathbb{G}_{n}(\beta, \beta') = 
P\left[
\mathbbm{1}\left\lbrace
Y\bX^{\T}\beta < 0
\right\rbrace \mathbbm{1}\left\lbrace
T_{n}(\bX) > \lambda_n
\right\rbrace
+ \mathbbm{1}\left\lbrace
Y\bX^{\T}\beta' < 0 
\right\rbrace \mathbbm{1}\left\lbrace
T_n(\bX) \le \lambda_n 
\right\rbrace
\right],
\end{equation*}
then it can be seen that $M(\beta^*) \triangleq  
\mathbb{G}_{n}(\beta^*, \beta^*)$.  Define 
$\widehat{\mathbb{G}}_{n}(\beta, \beta')$ to be the
estimator of $\mathbb{G}_{n}$ obtained by replacing 
$P$ with $\pn$.  The following result is the basis for
an adaptive projection interval
\begin{lem}\label{lemmyTheLemma}
Assume (A1)-(A4) and that $\lambda_n\to 0$ and $n\lambda_n\to\infty$ as $n\to\infty$. For any $\beta\in\mathbb{R}^p$  
then under either $P$ or $P_n$
\begin{equation*}
\sqrt{n}\left\lbrace
\widehat{\mathbb{G}}_{n}(\widehat{\beta}_{n}, \beta) 
- \mathbb{G}_{n}(\beta^*, \beta)
\right\rbrace \leadsto  \mathbb{W}(\beta^*, \beta),
\end{equation*}
where $\mathbb{W}(\beta^*, \beta)$ is normally distributed with mean zero
and variance $\rho^2(\beta^*,\beta)$.  
\end{lem}


A proof of this result, which includes a closed form expression for
$\rho^*(\beta^*, \beta)$, is given in the Appendix.  Let 
$\widehat{\rho}_{n}^2(\beta)$ denote a plugin estimator 
of $\rho^2(\beta^*, \beta)$ (see the Appendix for details). 
For any $\alpha \in (0,1)$ define 
\begin{equation*}
\mathfrak{W}_{n,1-\alpha}(\beta) = \left[
\widehat{\mathbb{G}}_{n}(\widehat{\beta}_{n}, \beta) - 
\frac{z_{1-\alpha/2}\widehat{\rho}_{n}(\beta)}{\sqrt{n}},\,
\widehat{\mathbb{G}}_{n}(\widehat{\beta}_{n}, \beta) + 
\frac{
z_{1-\alpha/2}\widehat{\rho}_{n}(\beta)
}{
\sqrt{n}
}
\right],
\end{equation*}
then it follows that 
$P_n\left\lbrace {\mathbb{G}}_{n}(\beta^*, \beta)
\in \mathfrak{W}_{n,1-\alpha}(\beta)\right\rbrace \ge 1-\alpha + o(1)$.  
For any $\omega, \alpha, \nu\in (0,1)$ such that $\omega = \alpha + \nu$, 
a $(1-\omega)\times 100\%$ adaptive projection confidence interval for 
$M(\beta^*,\beta^*)$ is 
  $\mathfrak{J}_{n,1-\omega} = \bigcup_{\beta \in \mathfrak{F}_{n,1-\eta}}
\mathfrak{W}_{n,1-\alpha}(\beta)$.  
While the form of the adaptive projection 
interval $\mathfrak{J}_{n,1-\omega}$ and 
the standard (non-adaptive) projection interval are similar,  
the union in $\mathfrak{J}_{n,1-\omega}$ only affects points
close to the decision boundary which can make it considerably less
conservative in some settings.  
The following result shows that
the adaptive projection interval provides asymptotically correct coverage;
a proof is provided in the Appendix.
\begin{thm}
Assume (A1), (A2), and (A4).  For each $n$ let 
$\left\lbrace (\bX_{i,n}, Y_{i,n}
)\right\rbrace_{i=1}^{n}$ comprise an $i.i.d.$ draw from $P_n$ which
satisfies (A3).  Let $M_n(\beta_n^*) = 
P_n \mathbbm{1}\left\lbrace Y\bX^{\T}\beta_n^* < 0\right\rbrace$  and
let $\omega \in (0, 1)$ be fixed. Then
$P_n\left\lbrace
M_n(\beta_n^*) \in \mathfrak{J}_{n,1-\omega} 
\right\rbrace \ge 1- \omega + o(1).$
\label{thm:adaptive}
\end{thm}
\noindent 
The preceding result was stated in terms of a normal-based confidence
interval for $\mathbb{G}_{n}(\beta^*, \beta)$ for each fixed $\beta$; 
an alternative, which may provide better finite sample performance,
is to use a bootstrap confidence interval  instead; extension of the
theory to handle this case is straightforward.

\subsubsection{Confidence intervals for $M(\widehat{c}_{n})$} 
To derive a confidence interval for $M(\widehat{\beta}_{n})$ we
take a similar approach to the adaptive projection interval in that
we partition the training into points that are near the decision 
boundary $\bx^{\T}\beta^* = 0$ and points that are far from
this boundary.  We take a sup over local perturbations of the
estimated optimal decision rule to construct an upper bound
on the generalization error and an inf to construct a lower bound;
we bootstrap these bounds to form a confidence set.  Because the
supremum (infimum) operation is smooth, the resultant bounds are regular 
and thereby can be consistently bootstrapped.  We derive an upper
bound with the derivation of a lower bound being analogous.  


Let $T_{n}(\bx)$ be as defined in the preceding section and write
\begin{eqnarray*}
\sqrt{n}\left\lbrace 
\widehat{M}_{n}(\widehat{\beta}_n) - M(\widehat{\beta}_{n}) 
\right\rbrace &=& \sqrt{n}\left(\pn-P\right)\mathbbm{1}
\left\lbrace
Y\bX^{\T}\widehat{\beta}_{n} < 0 
\right\rbrace\\
&=& 
\sqrt{n}(\pn - P)\mathbbm{1}\left\lbrace
Y\bX^{\T}\widehat{\beta}_{n} < 0
\right\rbrace
\mathbbm{1}\left\lbrace
T_{n}(\bX) > \lambda_n 
\right\rbrace \\ && \quad + \quad
\sqrt{n}\left(\pn-P\right)
\mathbbm{1}\left\lbrace  Y\bX^{\T}\widehat{\beta}_{n} < 0\right\rbrace
\mathbbm{1}\left\lbrace
T_{n}(\bX) \le \lambda_{n} 
\right\rbrace \\
&\le & 
\sqrt{n}(\pn - P)\mathbbm{1}\left\lbrace
Y\bX^{\T}\widehat{\beta}_{n} < 0
\right\rbrace
\mathbbm{1}\left\lbrace
T_{n}(\bX) > \lambda_n 
\right\rbrace \\ && \quad + \quad
\sup_{\beta\in\mathcal{S}_{n}}\sqrt{n}\left(\pn-P\right)
\mathbbm{1}\left\lbrace  Y\bX^{\T}\beta < 0\right\rbrace
\mathbbm{1}\left\lbrace
T_{n}(\bX) \le \lambda_{n} 
\right\rbrace, \\
&=& \mathcal{U}_{n}(\mathcal{S}_{n})
\end{eqnarray*} 
where $\mathcal{S}_{n}$ is any set that contains
$\widehat{\beta}_n$.  We define
$\mathcal{U}_n = \mathcal{U}_{n}(\mathbb{R}^p)$ which
we use to construct a confidence set; asymptotically, this
choice is unimprovable in some sense \citep[][]{laber2011adaptive}
though finite sample performance might be improved by
letting $\mathcal{S}_{n}$ depend on the data.  Define
a lower bound $\mathcal{L}_{n}$ analogously by replacing the $\sup$ with
an $\inf$ in the definition of $\mathcal{U}_{n}$.  

For any $\alpha \in (0,1)$ let $\widehat{\ell}_{n,\alpha/2}$ 
denote the $(\alpha/2)\times 100$ percentile of 
$\mathcal{L}_{n}$ and $\widehat{u}_{n,1-\alpha/2}$ denote
the $(1-\alpha/2)\times 100$ percentile of $\mathcal{U}_{n}$ then
\begin{equation}\label{coverage}
P\left\lbrace
\widehat{M}_{n}(\widehat{\beta}_n) - \widehat{u}_{n}/\sqrt{n} \le M(\widehat{\beta}_{n})
\le \widehat{M}_{n}(\widehat{\beta}_n) - \widehat{\ell}_{n}/\sqrt{n}
\right\rbrace \ge 1-\alpha.  
\end{equation}
Furthermore, if $\widehat{\ell}_{n}(\beta)$ 
is the $(\alpha/2)\times 100$ percentile of the conditional
distribution of $\mathcal{L}_{n}$ given $\widehat{\beta}_{n}=\beta$ 
and $\widehat{u}_{n}(\beta)$ is the $(1-\alpha/2)\times 100$
percentile of the conditional distribution of $\mathcal{U}_{n}$
given $\widehat{\beta}_{n}=\beta$ then 
 \begin{equation}\label{conditionalCoverage}
P\left\lbrace
\widehat{M}_{n}(\widehat{\beta}_n) - \widehat{u}_{n}(\widehat{\beta}_{n})/\sqrt{n} 
\le M(\widehat{\beta}_{n})
\le \widehat{M}_{n}(\widehat{\beta}_n) - \widehat{\ell}_{n}(\widehat{\beta}_{n})/\sqrt{n}
\Bigg| \widehat{\beta}_{n} \right\rbrace \ge 1-\alpha  
\end{equation}
with probability one. It can be verified directly that both
(\ref{coverage}) and (\ref{conditionalCoverage}) continue 
to hold under a sequence of local generative models
as in (A3).

In practice, the distributions of the bounds
$\mathcal{L}_n$ and $\mathcal{U}_n$ are not known and thus must
be estimated.  To estimate the unconditional distribution of 
these bounds we use the nonparametric bootstrap.  
Let $\mathbb{P}_{n}^{(b)}$ denote the bootstrap 
empirical distribution and for a functional
$f = f(P, \pn)$ write $f^{(b)} = f(\pn \mathbb{P}_{n}^{(b)})$ to
denote its bootstrap analog.  The following 
result is proved in \citet{laber2011adaptive}.
\begin{thm}
Assume (A1), (A2), and (A4) and let $\alpha \in (0,1)$ be fixed.  
Then 
\begin{equation*}
P\left\lbrace
\widehat{M}_{n}(\widehat{\beta}_n) - \widehat{u}_{n}^{(b)}/\sqrt{n}
\le M(\widehat{\beta}_{n}) \le
\widehat{M}_{n}(\widehat{\beta}_n) - \widehat{\ell}_{n}^{(b)}/\sqrt{n}
\right\rbrace  \ge 1-\alpha + o_P(1).  
\end{equation*}
\end{thm}
\noindent 
The inequality in the preceding result can be strengthened to equality
if $\bX^{\T}\beta^*$ is bounded away from zero with probability one.  

In order to construct a conditional confidence interval for
$M(\widehat{\beta}_{n})$ we first estimate the conditional
distribution of the bounds given $\widehat{\beta}_{n}$ and then
estimate their conditional quantiles.  While it is possible to derive
the joint asymptotic distribution of
$\left\lbrace \sqrt{n}\left(\widehat{\beta}_{n}-\beta^*\right),
  \mathcal{L}_{n}, \mathcal{U}_{n} \right\rbrace$,
the form of this limit is complex and hence it is difficult to extract
from it the requisite percentiles.  Instead, we use a kernel-based bootstrap
estimator.  Let $K$ denote a multivariate kernel function and $B$ a
matrix of bandwidth parameters.  Define the estimated conditional CDF
of $\mathcal{U}_{n}$ given $\widehat{\beta}_{n}$ as follows
 \begin{equation*}
 \widehat{F}_{\mathcal{U}_{n},n}\left(v\big|\beta\right) = 
 \frac{
 \mathbb{P}_{n}^{(b)}K\left\lbrace
 B^{-1}(\widehat{\beta}_{n}^{(b)}-\beta)
 \right\rbrace \mathbbm{1}\left\lbrace
 \mathcal{U}_{n}^{(b)} \le v
 \right\rbrace
 }{
 \mathbb{P}_{n}^{(b)}K\left\lbrace
 B^{-1}(\widehat{\beta}_{n}^{(b)}-\beta)
 \right\rbrace 
 },
 \end{equation*}
subsequently define $\widehat{u}_{n}(\beta) = 
\widehat{F}_{\mathcal{U}_{n},n}^{-1}
(1-\alpha/2;\widehat{\beta}_{n})$ to be the estimated
conditional $(1-\alpha/2)\times 100$ percentile 
of $\mathcal{U}_{n}$ given $\widehat{\beta}_{n}$.  
Define $\widehat{\ell}_{n}(\widehat{\beta}_{n})$ analogously. 
An asymptotic $(1-\alpha)\times 100\%$ conditional confidence set 
for $M(\widehat{\beta}_{n})$ is thus
\begin{equation*}
\left[
\widehat{M}_{n}(\widehat{\beta}_n) - \widehat{u}_{n}(\widehat{\beta}_{n})/\sqrt{n},
\widehat{M}_{n}(\widehat{\beta}_n) - \widehat{\ell}_{n}(\widehat{\beta}_{n})/\sqrt{n}
\right].
\end{equation*}
A rigorous analysis of the operating characteristics of this
interval is beyond the scope of this entry.  However, the proof
follows by showing: (i) the joint distribution of the bounds
and the sampling distribution of $\widehat{\beta}_{n}$ can
be consistently estimated using the bootstrap 
\citep[see][]{laber2011adaptive};
and (ii) that the conditional distribution can be consistently
estimated using the preceding kernel-based estimator
under an appropriate choice of bandwidth matrix
\citep[see][and references therein]{hall1999methods}.

\subsubsection{Confidence intervals for $M_n(\Gamma)$} 
Confidence intervals for $M_n(\Gamma)$ are less well-studied in 
the statistics literature primarily because they reflect
the average performance across repeated application of the same
learning algorithm in the same problem domain; however, it can be 
argued that methods like cross-validation that estimate the generalization
error by repeatedly splitting the data and averaging the performance
across splits are really estimating $M_n(\Gamma)$ rather than 
$M(\widehat{c}_{n})$ as is often purported 
\citep[see][for a discussion]{hastie2009elements}. We derive
one potential approach to constructing a confidence set
for $M_n(\Gamma)$ based on data-adaptive upper and lower bounds
as in the preceding section.

Let $\mathbb{F}_{n}$ denote the sampling distribution
of $\widehat{\beta}_{n}$, then
\begin{eqnarray*}
M_n(\Gamma) &=& \mathbb{E}P\mathbbm{1}\left\lbrace
Y\bX^{\T}\widehat{\beta}_{n} < 0 
\right\rbrace \\
&=& \int \mathbbm{1}\left\lbrace
y\bx^{\T}\beta < 0 
\right\rbrace dP(\bx, y)d\mathbb{F}_{n}(\beta) \\
&=& P \int \mathbbm{1}\left\lbrace 
Y\bX^{\T}\beta < 0 
\right\rbrace d\mathbb{F}_{n}(\beta),
\end{eqnarray*}
where the last equality follows from interchanging the order
of integration.  
Thus, if $\mathbb{F}_{n}^{(b)}$ is the bootstrap estimator
of $\mathbb{F}_{n}$ it can be seen that the plugin estimator
of $M_n(\Gamma)$ is 
\begin{eqnarray*}
\widehat{M}_{n}(\Gamma) &=&
\mathbb{E}_{M}\pn \mathbbm{1}\left\lbrace
Y\bX^{\T}\widehat{\beta}_{n}^{(b)} < 0
\right\rbrace
\\ 
&=&
\int \mathbbm{1}\left\lbrace
y\bx^{\T}\beta < 0 
\right\rbrace d\mathbb{P}_{n}(\bx, y)  d\mathbb{F}_{n}^{(b)}(\beta) \\
&=& \pn \int \mathbbm{1}\left\lbrace
Y\bX^{\T}\beta < 0 
\right\rbrace d\mathbb{F}_{n}^{(b)}(\beta),
\end{eqnarray*}
where $\mathbb{E}_{M}$ denotes expectation with respect to the
bootstrap resampling. 
Let $T_{n}(\bx)$ be as defined in the previous section. We will 
make use of the following bound
\begin{eqnarray*}
\sqrt{n}\left\lbrace
\widehat{M}_{n}(\Gamma) - M_n(\Gamma)
\right\rbrace &=& 
\sqrt{n}\left[
\pn \int \mathbbm{1}\left\lbrace
Y\bX^{\T}\beta < 0 
\right\rbrace d\mathbb{F}_{n}^{(b)}(\beta)
- 
P \int \mathbbm{1}\left\lbrace 
Y\bX^{\T}\beta < 0 
\right\rbrace d\mathbb{F}_{n}(\beta)
\right] \\
&\le &  \sqrt{n}\Bigg[
\pn \mathbbm{1} 
\left\lbrace 
T_{n}(\bX) > \lambda_{n} 
\right\rbrace 
\int 
\mathbbm{1}\left\lbrace
Y\bX^{\T}\beta < 0 
\right\rbrace d\mathbb{F}_{n}^{(b)}(\beta)
\\ 
&& \quad + \quad   
P 
\mathbbm{1} 
\left\lbrace 
T_{n}(\bX) > \lambda_{n} 
\right\rbrace
\int \mathbbm{1}\left\lbrace 
Y\bX^{\T}\beta < 0 
\right\rbrace d\mathbb{F}_{n}(\beta)
\Bigg] + \mathbb{S}_{n} \\
&=& \mathfrak{U}_{n},
\end{eqnarray*}
where 
\begin{multline*}
\mathbb{S}_{n} \triangleq  \sup_{\beta\in\mathbb{R}^p}
\sqrt{n}\Bigg[
\mathbb{E}_{M}\pn \mathbbm{1}\left\lbrace
T_{n}(\bX) \le \lambda_{n}
\right\rbrace 
\mathbbm{1}\left\lbrace
Y\bX^{\T}\sqrt{n}\left(
\widehat{\beta}_{n}^{(b)} - \beta^*
\right) + Y\bX^{\T}\beta < 0 
\right\rbrace
\\ +
\mathbb{E}P\mathbbm{1}\left\lbrace
T_{n}(\bX) \le \lambda_{n} 
\right\rbrace 
\mathbbm{1}\left\lbrace
Y\bX^{\T}\sqrt{n}(\widehat{\beta}_{n}-\beta^*) + Y\bX^{\T}\beta < 0 
\right\rbrace
\Bigg].  
\end{multline*}
To verify that $\mathfrak{U}_{n}$ is indeed an upper bound, one can
substitute $\sqrt{n}\beta^*$ in place of $\beta$.  
A lower bound, $\mathfrak{L}_{n}$, is obtained by replacing the $\sup$ 
with an $\inf$.  A confidence interval for $M_{n}(\Gamma)$ is 
obtained using the bootstrap distribution of the bounds
$\mathfrak{L}_{n}$ and $\mathfrak{U}_{n}$.  Let 
$\widehat{j}_{n,\alpha/2}$ denote the $(\alpha/2)\times 100$
percentile of $\mathfrak{L}_{n}^{(b)}$ and 
$\widehat{k}_{n,1-\alpha/2}$ the $(1-\alpha/2)\times 100$ percentile
of $\mathfrak{U}_{n}^{(b)}$.  The resultant approximate confidence 
interval for $M_{n}(\Gamma)$ is thus
\begin{equation*}
\left[
\widehat{M}_{n}(\Gamma) - \frac{\widehat{k}_{n,1-\alpha/2}}{\sqrt{n}},\,
\widehat{M}_{n}(\Gamma) - \frac{\widehat{j}_{n, \alpha/2}}{\sqrt{n}}
\right].
\end{equation*}
The operating characteristics of the above interval have not yet been
fully investigated; such an investigation is beyond the scope of this
entry, though we think this would make for interesting future work.

\section{Marginal mean outcome in decision making}
In a one-stage decision problem we assume that the observed data are
of the form $\left\lbrace (\bX_i, A_i, Y_i)\right\rbrace_{i=1}^{n}$
which comprise $n$ independent replicates of the triple $(\bX, A, Y)$,
where: $\bX\in\mathbb{R}^p$ describes the decision context;
$A\in\left\lbrace -1,1\right\rbrace$ is the decision; and
$Y\in\mathbb{R}$ is the outcome coded so that higher values are
better.  A common application is precision medicine wherein:
$\bX$ denotes baseline
patient characteristics, $A$ is their assigned treatment, and $Y$ is
the patient's clinical outcome 
\citep[][]{mintron, yingqi, baqun, bibhasBook, kosorok2015adaptive}.  A
decision rule, $d:\mathrm{dom}\,\bX \rightarrow \mathrm{dom}\,A$, so
that a decision maker following $d$ would make decision $d(\bx)$ in
the context $\bX=\bx$.  An optimal decision rules, $d^{\mathrm{opt}}$,
maximized the mean outcome if applied over the distribution of
contexts.  The one-stage decision problem is closely related to the
classification problem studied previously.  The critical difference
lies in the information contained in the available data.  In a
decision problem as described above, the outcome $Y$ provides indirect
feedback about the quality of decision $A$ in context $\bX$; in
contrast, in a classification problem, one would be given the context
$\bX$ {\em and} the optimal decision $d^{\mathrm{opt}}(\bX)$
\citep[see][]{sutton}.

 We formalize an optimal regime using the language of potential outcomes
 \cite[][]{rubin,splawa1990application}.  Let 
 $Y^*(a)$   denote the potential outcome
 under decision $a\in\mathrm{dom}\,A$. Then the potential outcome
 under a decision rule $d$ is 
 $Y^*(d) = \sum_{a}Y^*(a)\mathbbm{1}\left\lbrace d(\bX) =a \right\rbrace$.  
 Given a class of decision rules, $\mathcal{D}$, we define an optimal
 decision rule, $d^{\mathrm{opt}}\in\mathcal{D}$, as one that satisfies
 $PY^*(d^{\mathrm{opt}}) \ge PY^*(d)$ for all $d\in\mathcal{D}$.  
 In order to identify $d^{\mathrm{opt}}$ in terms of the data-generating
 model, we make the following assumptions: (C1) positivity,
 $\pi(a;\bx) \triangleq P(A=a|\bX=\bx) > 0$ with probability one for each 
 $a\in\mathrm{dom}\,A$ and $\bx\in\mathrm{dom}\,\bX$; 
 (C2) no unmeasured confounding, 
 $\left\lbrace Y^*(a)\,:\,a\in\mathrm{dom}\,A\right\rbrace \perp
 A\big| \bX$; and (C3) consistency, $Y = Y^*(A)$.  These
 assumptions are standard in single stage
 decision problems \citep[][]{baqun}; 
 (C1) and (C2) can be guaranteed in a randomized
 experiment but (C2) is  not verifiable in an observational 
 studies.  

 Under (C1)-(C3) it follows that the marginal mean outcome under a given
 decision rule $d$ can be expressed as 
 \begin{equation}\label{ipw}
 V(d) = P\left[
 \frac{
 Y\mathbbm{1}\left\lbrace d(\bX) = A \right\rbrace
 }{
 P\left(A\big|\bX\right)
 }
 \right]
= P\left[
 \frac{
 Y\mathbbm{1}\left\lbrace d(\bX) \ne -A \right\rbrace
 }{
 P\left(A\big|\bX\right)
 }
 \right]
,
 \end{equation}
which can be viewed as a weighted misclassification rate for 
$d$ with labels $-A$ and weights $Y/P(A|\bX)$; however, in this 
formulation it should be emphasized that all the unobserved actions
$-A$ need not be the correct labels, i.e., the need need not equal
$d^{\mathrm{opt}}(\bX)$, indeed if there observed data 
were generated in a randomized clinical trial $A$ would be 
assigned at random 
\cite[see][for additional discussion]{zhang2012estimating,
baqun2,zhang2015c}.  Nevertheless,  
the expression for $V(d)$ in (\ref{ipw}) can be used to directly
extend the inferential methods for the generalization error in
classification to the marginal mean outcome in the 
single-stage decision setting. 

As in the classification setting, for the purpose of illustration, we
consider linear models fit using least squares; extensions to other
smooth parametric models and convex loss functions is straightforward. 
Define $Q(\bx, a) = \mathbb{E}(Y|\bX=\bx, A=a)$ then, under (C1)-(C3), 
$d^{\mathrm{opt}}(\bx) = \arg\max_{a}Q(\bx, a)$ 
\citep[][]{murphyZFive,mintron, schulte, kosorok2015adaptive}. 
Thus, a natural approach
to estimating $d^{\mathrm{opt}}$ based on the preceding characterization 
is to construct an estimator
 $\widehat{Q}_{n}(\bx, a)$ of $Q(\bx,a)$ and then to use
the plugin estimator $\widehat{d}_{n}(\bx) = 
\arg\max_{a}\widehat{Q}_{n}(\bx, a)$.   We take this approach
using linear working models of the form
$Q(\bx, a;\beta) = \bx_{0}^{\T}\beta_{0} + a\bx_{1}^{\T}\beta_1$,
where $\beta = (\beta_0^{\T}, \beta_1^{\T})^{\T}$ and
$\bx_0, \bx_1$ are features constructed from $\bx$.  
Define $\widehat{\beta}_{n} = \arg\min_{\beta}
\pn \left\lbrace Y - Q(\bX, A;\beta)\right\rbrace^2$ and 
subsequently $\widehat{d}_{n}(\bx) = \arg\max_a 
Q(\bx, a;\widehat{\beta}_{n})$.  Define
the population analog of $\widehat{\beta}_{n}$ to
be $\beta^* = \arg\min_{\beta}P\left\lbrace Y -Q(\bX, A;\beta)
\right\rbrace^2$ and subsequently $d^{\mathrm{opt}}(\bx) =
\arg\max_{a}Q(\bx, a;\beta^*)$.\footnote{
Note that $d^{\mathrm{opt}}$ need not be the optimal
decision rule among all linear decision rules even if the
optimal rule over the space of measurable rules is 
linear \citep[][]{mintron}.
}

Given $\beta$, write $V(\beta)$ to denote the expected outcome
under the rule $d(\bx) =\arg\max_a Q(\bx, a;\beta) = \mathrm{sign}
\left(
\bx_{1}^{\T}\beta_1
\right)$.  Thus,
$V(\beta^*)$ denotes the marginal mean outcome associated
with the decision rule $d^{\mathrm{opt}}$; thus, $V(\beta^*)$
is analogous to $M(\beta^*)$ and one can use a projection
interval to construct an asymptotically valid confidence set.   
Suppose first that the propensity 
score, $\pi(a;\bx)$ is known, and for fixed $\beta$ define
$\widehat{V}_{n}(\beta) = 
\pn\left[
Y\mathbbm{1}\left\lbrace
-A\bX_{1}^{\T}\beta_1 < 0
\right\rbrace/\pi(A; \bX)
\right]
$.  For any fixed $\beta$ it follows from the central limit theorem
that $\sqrt{n}\left\lbrace \widehat{V}_{n}(\beta) - V(\beta)\right\rbrace$
is asymptotically normal with mean zero and variance 
\begin{equation*}
\varsigma^{*2}(\beta) = 
P\left[
\frac{
Y\mathbbm{1}\left\lbrace
-A\bX_1^{\T}\beta_1 < 0 
\right\rbrace
}{
\pi(\bX;A)
}
- P\left\lbrace
\frac{
Y\mathbbm{1}\left\lbrace
-A\bX_1^{\T}\beta_1 < 0 
\right\rbrace
}{
\pi(\bX;A)
}
\right\rbrace
\right]^2.
\end{equation*}
Let $\widehat{\varsigma}_{n}^2(\beta)$ denote the plug-in estimator
of $\varsigma^{*2}(\beta)$ obtained by replacing $P$ with $\pn$ in the
foregoing expression. For any $\alpha \in (0,1)$, define
\begin{equation*}
\mathfrak{V}_{n, 1-\alpha}(\beta) \triangleq 
\left[
\widehat{V}_{n}(\beta) - 
\frac{
z_{1-\alpha/2}
\widehat{\varsigma}_{n}(\beta)
}{
\sqrt{n}
},\,
\widehat{V}_{n}(\beta) + 
\frac{
z_{1-\alpha/2}
\widehat{\varsigma}_{n}(\beta)
}{
\sqrt{n}
}
\right],
\end{equation*} 
which is an asymptotic $(1-\alpha)\times 100\%$ confidence interval for
$V(\beta)$.  Given $\eta \in (0,1)$ let $\mathfrak{Q}_{n,1-\eta}$ be a
$(1-\eta)\times 100\%$ confidence set for $\beta^*$ (e.g., one could
construct a Wald-type confidence set as in the preceding section).
For any $\omega \in (0,1)$ choose $\eta, \alpha \in (0,1)$ so that
$\omega = \alpha + \eta$, then a $(1-\omega)\times 100\%$ confidence
set for $V(\beta^*)$ is
$\bigcup_{\beta\in\mathfrak{Q}_{n,1-\eta}}
\mathfrak{V}_{n,1-\alpha}(\beta)$.
The theoretical properties of the projection interval for $M(\beta^*)$
developed in the preceding section port directly to the confidence
interval for $V(\beta^*)$; moreover, one can also construct an
adaptive projection interval to reduce conservatism.  The extension to
the case where the propensity score is unknown is straightforward
provided that one estimates the propensity score using a regular
asymptotically linear estimator.  

The marginal mean outcome for the estimated optimal decision rule
$\widehat{d}_{n}$ is $V(\widehat{\beta}_{n}) = 
P\left[Y\mathbbm{1}\left\lbrace -A\bX_1^{\T}\widehat{\beta}_{1,n} < 0
\right\rbrace /\pi(A;\bX)\right]$ which is analogous to
$M_n(\widehat{\beta}_{n})$ in the classification case.  
Define $\widehat{V}_{n} =    
\pn \left[Y\mathbbm{1}\left\lbrace -A\bX_1^{\T}\widehat{\beta}_{1,n} < 0
\right\rbrace /\pi(A;\bX)\right]$, mimicking the classification setting,
we construct confidence bounds for $V(\widehat{\beta}_{n})$ 
using smooth upper and lower bounds
on $\sqrt{n}\left\lbrace \widehat{V}_{n} - V(\widehat{\beta}_{n})
\right\rbrace$; the derivation of these bounds is essentially
identical to those used in the classification setting.
Let $\widehat{R}_{n}$ denote the plug-in estimator
of the asymtptotic covariance of $\widehat{\beta}_{1,n}$ and
define $Z_n(\bX) = 
n(\bX_1^{\T}\widehat{\beta}_{1,n})^2/\bX_1^{\T}\widehat{R}_{n}
\bX_1$.  For any sequence of non-negative tuning parameters 
$\left\lbrace \varrho_{n}\right\rbrace_{n\ge 1}$ the following
inequality holds
\begin{multline*}
\sqrt{n}\left\lbrace
\widehat{V}_{n}- V(\widehat{\beta}_{n})
\right\rbrace \le  \sqrt{n}\left(
\pn - P
\right)\left[
\frac{
Y\mathbbm{1}\left\lbrace
-A\bX_1^{\T}\widehat{\beta}_{1,n} < 0
\right\rbrace \mathbbm{1}\left\lbrace Z_n(\bX) > \varrho_n\right\rbrace
}{
\pi(A;\bX)
}
\right] 
\\ + 
\sup_{\beta\in\mathcal{J}_{n}}
\sqrt{n}\left(
\pn - P
\right)\left[
\frac{
Y\mathbbm{1}\left\lbrace
-A\bX_1^{\T}\beta < 0
\right\rbrace \mathbbm{1}\left\lbrace Z_n(\bX) \le \varrho_n\right\rbrace
}{
\pi(A;\bX)
}
\right],
\end{multline*}
provided that $\widehat{\beta}_{1,n} \in \mathcal{J}_{n}$.  
Let $\mathcal{E}_{n}$ denote the upper bound obtained by setting
$\mathcal{J}_{n} = \mathbb{R}^{\dim\,\bX_1}$ and let 
$\mathcal{M}_{n}$ denote an analogous lower bound obtained by 
replacing the $\sup$ with an $\inf$ in the definition of $\mathcal{E}_n$. 
A confidence interval for $V(\widehat{\beta}_{n})$ is obtained
 by taking the percentiles of the bootstrap distribution of
these upper and lower bounds; theoretical properties can be derived
from analogous arguments to those used in for bound-based inference
for $M(\widehat{\beta}_{n})$.  Similarly, one can derive bounds 
for $\mathbb{E}V(\widehat{\beta}_{n})$ my mimicking those
for $M_{n}(\Gamma)$; thus, we omit such derivations.

\section{Discussion}
We provided a whirlwind tour of confidence intervals for the 
generalization error in classification and the marginal mean 
outcome in single-stage decision problems.  We delineated three
types of generalization error and argued that: (i) these need not
be close even asymptotically; and (ii) even when they converge to the same
limit this convergence is not uniform and consequently they need not
be close in finite samples.  The implication is that inference
procedures for the generalization error (or the marginal mean outcome
in decision problems) must be consistent under moving parameter
asymptotics if they are to be relied upon in practice.  

There
are a number of interesting extensions and alternative approaches
that we did not cover in this entry.  Perhaps the biggest omission was a 
lack
of discussion of subsampling-based approaches. The m-out-of-n bootstrap
and substampling without replacement are a common approach to inference
in nonregular problems like those considered here
\citep[][]{chakraborty2014inference, mofn}.  
However, such
procedures can be difficult to tune and in some cases perform
worse than naive bootstrap or normal approximations 
\citep[][]{samworth2003note}.   
The inference procedures
we reviewed (and the few new ones we introduced) were based on asymptotic
arguments.  However, in some settings, finite sample (i.e., non-asymptotic)
bounds may be desired.  There is a large body of literature on such
bounds in statistics and computer science 
\citep[see][and references therein]{murphyZFive, mintron}.  These
bounds can provide a more refined characterization of the relationship
between the instability of the generalization error in finite samples
and the amassing of points near the optimal decision boundary.  In
the context of decision problems an important extension is to multi-stage
settings; the primary complication in this extension is that the
estimators indexing the underlying models are nonregular so that
bound-based and projection-based intervals become considerably more
involved \citep[][]{laber2014statistical}.

\section{Acknowledgments}
This work was partially supported by NSF grantsDMS-1555141 and
DMS-1513579 and NIH grants R01-DK-108073, 1R01 AA023187-01A1, P01 CA142538, and R21-MH-108999.  

\section{Related articles}
Davidian, Marie, Anastasios A. Tsiatis, and Eric B. Laber. "Optimal Dynamic Treatment Regimes." Wiley StatsRef: Statistics Reference Online (2016).

\nocite{van1998asymptotic} 
\bibliographystyle{Chicago}
\bibliography{sampleSize.bib}

\begin{thebibliography}{}

\bibitem[\protect\citeauthoryear{Amari, Fujita, and Shinomoto}{Amari
  et~al.}{1992}]{amari1992four}
Amari, S.-i., N.~Fujita, and S.~Shinomoto (1992).
\newblock Four types of learning curves.
\newblock {\em Neural Computation\/}~{\em 4\/}(4), 605--618.

\bibitem[\protect\citeauthoryear{Bartlett, Jordan, and McAuliffe}{Bartlett
  et~al.}{2006}]{bartlett2006convexity}
Bartlett, P.~L., M.~I. Jordan, and J.~D. McAuliffe (2006).
\newblock Convexity, classification, and risk bounds.
\newblock {\em Journal of the American Statistical Association\/}~{\em
  101\/}(473), 138--156.

\bibitem[\protect\citeauthoryear{Berger and Boos}{Berger and
  Boos}{1994}]{berger1994p}
Berger, R.~L. and D.~D. Boos (1994).
\newblock P values maximized over a confidence set for the nuisance parameter.
\newblock {\em Journal of the American Statistical Association\/}~{\em
  89\/}(427), 1012--1016.

\bibitem[\protect\citeauthoryear{Casella}{Casella}{1992}]{casella1992conditional}
Casella, G. (1992).
\newblock Conditional inference from confidence sets.
\newblock {\em Lecture Notes-Monograph Series\/}, 1--12.

\bibitem[\protect\citeauthoryear{Chakraborty, Laber, and Zhao}{Chakraborty
  et~al.}{2013}]{mofn}
Chakraborty, B., E.~B. Laber, and Y.~Zhao (2013).
\newblock Inference for optimal dynamic treatment regimes using an adaptive
  m-out-of-n bootstrap scheme.
\newblock {\em Biometrics\/}~{\em 69\/}(3), 714--723.

\bibitem[\protect\citeauthoryear{Chakraborty, Laber, and Zhao}{Chakraborty
  et~al.}{2014}]{chakraborty2014inference}
Chakraborty, B., E.~B. Laber, and Y.-Q. Zhao (2014).
\newblock Inference about the expected performance of a data-driven dynamic
  treatment regime.
\newblock {\em Clinical Trials\/}~{\em 11\/}(4), 408--417.

\bibitem[\protect\citeauthoryear{Chakraborty and Moodie}{Chakraborty and
  Moodie}{2013}]{bibhasBook}
Chakraborty, B. and E.~E. Moodie (2013).
\newblock {\em Statistical Methods for Dynamic Treatment Regimes}.
\newblock Springer.

\bibitem[\protect\citeauthoryear{Dawid}{Dawid}{1994}]{dawid1994selection}
Dawid, A. (1994).
\newblock Selection paradoxes of bayesian inference.
\newblock {\em Lecture Notes-Monograph Series\/}, 211--220.

\bibitem[\protect\citeauthoryear{Duda, Hart, and Stork}{Duda
  et~al.}{2012}]{duda2012pattern}
Duda, R.~O., P.~E. Hart, and D.~G. Stork (2012).
\newblock {\em Pattern classification}.
\newblock John Wiley \& Sons.

\bibitem[\protect\citeauthoryear{Efron and Tibshirani}{Efron and
  Tibshirani}{1997}]{efron1997improvements}
Efron, B. and R.~Tibshirani (1997).
\newblock Improvements on cross-validation: the 632+ bootstrap method.
\newblock {\em Journal of the American Statistical Association\/}~{\em
  92\/}(438), 548--560.

\bibitem[\protect\citeauthoryear{Efron and Tibshirani}{Efron and
  Tibshirani}{1994}]{efron1994introduction}
Efron, B. and R.~J. Tibshirani (1994).
\newblock {\em An introduction to the bootstrap}, Volume~57.
\newblock CRC press.

\bibitem[\protect\citeauthoryear{Hall, Wolff, and Yao}{Hall
  et~al.}{1999}]{hall1999methods}
Hall, P., R.~C. Wolff, and Q.~Yao (1999).
\newblock Methods for estimating a conditional distribution function.
\newblock {\em Journal of the American Statistical association\/}~{\em
  94\/}(445), 154--163.

\bibitem[\protect\citeauthoryear{Hastie, Tibshirani, Friedman, Hastie,
  Friedman, and Tibshirani}{Hastie et~al.}{2009}]{hastie2009elements}
Hastie, T., R.~Tibshirani, J.~Friedman, T.~Hastie, J.~Friedman, and
  R.~Tibshirani (2009).
\newblock {\em The elements of statistical learning}, Volume~2.
\newblock Springer.

\bibitem[\protect\citeauthoryear{Haussler, Kearns, Seung, and Tishby}{Haussler
  et~al.}{1996}]{haussler1996rigorous}
Haussler, D., M.~Kearns, H.~S. Seung, and N.~Tishby (1996).
\newblock Rigorous learning curve bounds from statistical mechanics.
\newblock {\em Machine Learning\/}~{\em 25\/}(2-3), 195--236.

\bibitem[\protect\citeauthoryear{Hirano and Porter}{Hirano and
  Porter}{2012}]{hirano2012}
Hirano, K. and J.~R. Porter (2012).
\newblock Impossibility results for nondifferentiable functionals.
\newblock {\em Econometrica\/}~{\em 80\/}(4), 1769--1790.

\bibitem[\protect\citeauthoryear{Kosorok and Moodie}{Kosorok and
  Moodie}{2015}]{kosorok2015adaptive}
Kosorok, M.~R. and E.~E. Moodie (2015).
\newblock {\em Adaptive treatment strategies in practice: planning trials and
  analyzing data for personalized medicine}.
\newblock SIAM.

\bibitem[\protect\citeauthoryear{Laber, Lizotte, Qian, Pelham, and
  Murphy}{Laber et~al.}{2014}]{laber2014statistical}
Laber, E.~B., D.~J. Lizotte, M.~Qian, W.~E. Pelham, and S.~A. Murphy (2014).
\newblock Dynamic treatment regimes: Technical challenges and applications.
\newblock {\em Electronic journal of statistics\/}~{\em 8\/}(1), 1225.

\bibitem[\protect\citeauthoryear{Laber and Murphy}{Laber and
  Murphy}{2011}]{laber2011adaptive}
Laber, E.~B. and S.~A. Murphy (2011).
\newblock Adaptive confidence intervals for the test error in classification.
\newblock {\em Journal of the American Statistical Association\/}~{\em
  106\/}(495), 904--913.

\bibitem[\protect\citeauthoryear{Laber, Shedden, and Yang}{Laber
  et~al.}{2016}]{laber2016imputation}
Laber, E.~B., K.~Shedden, and Y.~Yang (2016).
\newblock An imputation method for estimating the learning curve in
  classification problems.
\newblock In {\em Statistical Analysis for High-Dimensional Data}, pp.\
  189--209. Springer.

\bibitem[\protect\citeauthoryear{Laber, Zhao, Regh, Davidian, Tsiatis,
  Stanford, Zeng, Song, and Kosorok}{Laber et~al.}{2016}]{laber2016using}
Laber, E.~B., Y.-Q. Zhao, T.~Regh, M.~Davidian, A.~Tsiatis, J.~B. Stanford,
  D.~Zeng, R.~Song, and M.~R. Kosorok (2016).
\newblock Using pilot data to size a two-arm randomized trial to find a nearly
  optimal personalized treatment strategy.
\newblock {\em Statistics in medicine\/}~{\em 35\/}(8), 1245--1256.

\bibitem[\protect\citeauthoryear{Luckett, Laber, El-Kamary, Fan, Jhaveri,
  Perou, Shebl, and Kosorok}{Luckett et~al.}{2018}]{luckett2018ROC}
Luckett, D., E.~Laber, S.~El-Kamary, C.~Fan, R.~Jhaveri, C.~Perou, F.~Shebl,
  and M.~Kosorok (2018).
\newblock Receiver operating characteristic curves and confidence bands for
  support vector machines.
\newblock {\em Under review\/}.

\bibitem[\protect\citeauthoryear{Mukherjee, Tamayo, Rogers, Rifkin, Engle,
  Campbell, Golub, and Mesirov}{Mukherjee
  et~al.}{2003}]{mukherjee2003estimating}
Mukherjee, S., P.~Tamayo, S.~Rogers, R.~Rifkin, A.~Engle, C.~Campbell, T.~R.
  Golub, and J.~P. Mesirov (2003).
\newblock Estimating dataset size requirements for classifying dna microarray
  data.
\newblock {\em Journal of computational biology\/}~{\em 10\/}(2), 119--142.

\bibitem[\protect\citeauthoryear{Murphy}{Murphy}{2005}]{murphyZFive}
Murphy, S.~A. (2005, Jul).
\newblock A generalization error for {Q}-learning.
\newblock {\em Journal of Machine Learning Research\/}~{\em 6}, 1073--1097.

\bibitem[\protect\citeauthoryear{Qian and Murphy}{Qian and
  Murphy}{2011}]{mintron}
Qian, M. and S.~Murphy (2011).
\newblock {Performance Guarantees for Individualized Treatment Rules}.
\newblock {\em The Annals of Statistics\/}~{\em 39\/}(2), 1180--1210.

\bibitem[\protect\citeauthoryear{Robins, Rotnitzky, et~al.}{Robins
  et~al.}{2014}]{robins2014discussion}
Robins, J., A.~Rotnitzky, et~al. (2014).
\newblock Discussion of “dynamic treatment regimes: Technical challenges and
  applications”.
\newblock {\em Electronic Journal of Statistics\/}~{\em 8\/}(1), 1273--1289.

\bibitem[\protect\citeauthoryear{Robins}{Robins}{2004}]{robinsTF}
Robins, J.~M. (2004).
\newblock Optimal structural nested models for optimal sequential decisions.
\newblock In {\em Proceedings of the Second Seattle Symposium on Biostatitics},
  pp.\  189--326. Springer.

\bibitem[\protect\citeauthoryear{Rubin}{Rubin}{1978}]{rubin}
Rubin, D. (1978).
\newblock Bayesian inference for causal effects: The role of randomization.
\newblock {\em The Annals of Statistics\/}~{\em 6\/}(1), 34--58.

\bibitem[\protect\citeauthoryear{Samworth}{Samworth}{2003}]{samworth2003note}
Samworth, R. (2003).
\newblock A note on methods of restoring consistency to the bootstrap.
\newblock {\em Biometrika\/}~{\em 90\/}(4), 985--990.

\bibitem[\protect\citeauthoryear{Schiavo and Hand}{Schiavo and
  Hand}{2000}]{schiavo2000ten}
Schiavo, R.~A. and D.~J. Hand (2000).
\newblock Ten more years of error rate research.
\newblock {\em International Statistical Review\/}~{\em 68\/}(3), 295--310.

\bibitem[\protect\citeauthoryear{Schulte, Tsiatis, Laber, , and
  Davidian}{Schulte et~al.}{2014}]{schulte}
Schulte, P., A.~Tsiatis, E.~Laber, , and M.~Davidian (2014).
\newblock Q- and a-learning methods for estimating optimal dynamic treatment
  regimes.
\newblock {\em Statistical Science\/}~{\em 29\/}(4), 640--661.

\bibitem[\protect\citeauthoryear{Seber and Lee}{Seber and
  Lee}{2012}]{seber2012linear}
Seber, G.~A. and A.~J. Lee (2012).
\newblock {\em Linear regression analysis}, Volume 936.
\newblock John Wiley \& Sons.

\bibitem[\protect\citeauthoryear{Shao and Tu}{Shao and
  Tu}{2012}]{shao2012jackknife}
Shao, J. and D.~Tu (2012).
\newblock {\em The jackknife and bootstrap}.
\newblock Springer Science \& Business Media.

\bibitem[\protect\citeauthoryear{Splawa-Neyman, Dabrowska, Speed,
  et~al.}{Splawa-Neyman et~al.}{1990}]{splawa1990application}
Splawa-Neyman, J., D.~Dabrowska, T.~Speed, et~al. (1990).
\newblock On the application of probability theory to agricultural experiments.
  essay on principles. section 9.
\newblock {\em Statistical Science\/}~{\em 5\/}(4), 465--472.

\bibitem[\protect\citeauthoryear{Stefanski and Boos}{Stefanski and
  Boos}{2002}]{stefanski2002calculus}
Stefanski, L.~A. and D.~D. Boos (2002).
\newblock The calculus of m-estimation.
\newblock {\em The American Statistician\/}~{\em 56\/}(1), 29--38.

\bibitem[\protect\citeauthoryear{Sutton and Barto}{Sutton and
  Barto}{1998}]{sutton}
Sutton, R. and A.~Barto (1998).
\newblock {\em Reinforcment Learning: An Introduction}.
\newblock The MIT Press.

\bibitem[\protect\citeauthoryear{Tsiatis}{Tsiatis}{2007}]{tsiatis2007semiparametric}
Tsiatis, A. (2007).
\newblock {\em Semiparametric theory and missing data}.
\newblock Springer Science \& Business Media.

\bibitem[\protect\citeauthoryear{Van~der Vaart}{Van~der
  Vaart}{1991}]{van1991differentiable}
Van~der Vaart, A. (1991).
\newblock On differentiable functionals.
\newblock {\em The Annals of Statistics\/}, 178--204.

\bibitem[\protect\citeauthoryear{Van~der Vaart}{Van~der
  Vaart}{1998}]{van1998asymptotic}
Van~der Vaart, A.~W. (1998).
\newblock {\em Asymptotic statistics}, Volume~2.
\newblock Cambridge university press.

\bibitem[\protect\citeauthoryear{Zhang, Tsiatis, Davidian, Zhang, and
  Laber}{Zhang et~al.}{2012}]{zhang2012estimating}
Zhang, B., A.~A. Tsiatis, M.~Davidian, M.~Zhang, and E.~Laber (2012).
\newblock Estimating optimal treatment regimes from a classification
  perspective.
\newblock {\em Stat\/}~{\em 1\/}(1), 103--114.

\bibitem[\protect\citeauthoryear{Zhang, Tsiatis, Laber, and Davidian}{Zhang
  et~al.}{2012}]{baqun}
Zhang, B., A.~A. Tsiatis, E.~B. Laber, and M.~Davidian (2012).
\newblock A robust method for estimating optimal treatment regimes.
\newblock {\em Biometrics\/}~{\em 68\/}(4), 1010--1018.

\bibitem[\protect\citeauthoryear{Zhang, Tsiatis, Laber, and Davidian}{Zhang
  et~al.}{2013}]{baqun2}
Zhang, B., A.~A. Tsiatis, E.~B. Laber, and M.~Davidian (2013).
\newblock Robust estimation of optimal dynamic treatment regimes for sequential
  treatment decisions.
\newblock {\em Biometrika\/}~{\em 100\/}(3), 681--694.

\bibitem[\protect\citeauthoryear{Zhang and Zhang}{Zhang and
  Zhang}{2015}]{zhang2015c}
Zhang, B. and M.~Zhang (2015).
\newblock C-learning: A new classification framework to estimate optimal
  dynamic treatment regimes.
\newblock {\em Biometrics\/}.

\bibitem[\protect\citeauthoryear{Zhang}{Zhang}{1995}]{zhang1995ape}
Zhang, P. (1995).
\newblock Ape and models for categorical panel data.
\newblock {\em Scandinavian Journal of Statistics\/}~{\em 22}, 83--94.

\bibitem[\protect\citeauthoryear{Zhao, Zeng, Rush, and Kosorok}{Zhao
  et~al.}{2012}]{yingqi}
Zhao, Y., D.~Zeng, A.~J. Rush, and M.~R. Kosorok (2012).
\newblock Estimating individualized treatment rules using outcome weighted
  learning.
\newblock {\em Journal of the American Statistical Association\/}~{\em
  107\/}(499), 1106--1118.

\end{thebibliography}

\section{Appendix: additional technical details} 
\subsection{Proof of Lemma \ref{lemmyTheLemma}}
We can rewrite $\sqrt n(\widehat{\mathbb{G}}_n(\widehat{\beta}_n,\beta)-
\mathbb{G}_n(\beta_n^*,\beta)) $ as
\begin{eqnarray}
&&\sqrt n(\widehat{\mathbb{G}}_n(\widehat{\beta}_n,\beta)-
\mathbb{G}_n(\beta_n^*,\beta)) \nonumber \\
& = & 
P_n\left[\sqrt n
\left(\mathbbm{1}\left\lbrace
Y\bX^{\T}\widehat\beta_n < 0
\right\rbrace  - \mathbbm{1}\left\lbrace
Y\bX^{\T}\beta_n^* < 0
\right\rbrace\right)
\mathbbm{1}\left\lbrace
T_{n}(\bX) > \lambda_n
\right\rbrace
\right] \nonumber\\
&& + \sqrt n(\mathbb{P}_n-P_n)\left[
\left(\mathbbm{1}\left\lbrace
Y\bX^{\T}\widehat\beta_n < 0
\right\rbrace  - \mathbbm{1}\left\lbrace
Y\bX^{\T}\beta_n^* < 0
\right\rbrace\right)
\mathbbm{1}\left\lbrace
T_{n}(\bX) > \lambda_n
\right\rbrace
\right] \nonumber\\
&& + \sqrt n (\mathbb{P}_n-P_n)\left[
\mathbbm{1}\left\lbrace
Y\bX^{\T}\beta_n^* < 0
\right\rbrace \mathbbm{1}\left\lbrace
T_{n}(\bX) > \lambda_n
\right\rbrace
+ \mathbbm{1}\left\lbrace
Y\bX^{\T}\beta < 0 
\right\rbrace \mathbbm{1}\left\lbrace
T_n(\bX) \le \lambda_n 
\right\rbrace
\right] \label{eqn: lemprof1}
\end{eqnarray}
Under Assumptions (A1)-(A3), $\sqrt n (\widehat{\beta}_n-\beta_n^*)\stackrel{d}{\to} N(0, \Sigma(\beta^*))$ and $\sqrt n(\beta^*_n-\beta^*) = O(1)$. Thus for every $\bx\in\mathbb{R}^p$, 
$$nT_n(\bx) = n[\bx^\T(\widehat{\beta}_n-\beta_n^*)
+ \bx^\T({\beta}^*_n-\beta^*)+ \bx^\T\beta^*]^2/(\bx^\T\widehat{\Sigma}_n\bx)
= [\bx^\T(\sqrt n \beta^* + O_P(1))]^2/(\bx^\T\widehat{\Sigma}_n\bx).$$
By Continuous Mapping Theorem, under Assumption (A4) and the condition that $\lambda_n \to 0$ and $n\lambda_n\to\infty$,
 $\mathbbm{1}\left\lbrace
T_n(\bx) > \lambda_n 
\right\rbrace \to \mathbbm{1}\left\lbrace
\bx^\T\beta^*\neq 0
\right\rbrace$ in probability pointwise. Similarly, we can show that
$\sqrt n
	\left(\mathbbm{1}\left\lbrace
	y\bx^{\T}\widehat\beta_n < 0
	\right\rbrace  - \mathbbm{1}\left\lbrace
	y\bx^{\T}\beta_n^* < 0
	\right\rbrace\right)
	\mathbbm{1}\left\lbrace
	T_{n}(\bx) > \lambda_n
	\right\rbrace$
converges pointwise to zero in probability. Using Dominated Convergence Theorem, we have
$$P\left[ |\mathbbm{1}\left\lbrace
T_{n}(\bX) > \lambda_n
\right\rbrace - \mathbbm{1}\left\lbrace
\bX^\T\beta^*\neq 0
\right|\right] \to 0 
$$ in probability, and the first term of (\ref{eqn: lemprof1}) converges to zero in probability.
Using similar arguments as those in the proof of Lemma 19.24 of Van der Vaart (1998), the second term of (\ref{eqn: lemprof1}) converges to zero in probability, and the third term converges to a normal distribution with mean zero and variance
\begin{equation*}
\rho^2(\beta^*,\beta) = {\rm Var}\left(\mathbbm{1}\left\lbrace
Y\bX^{\T}\beta^* < 0
\right\rbrace \mathbbm{1}\left\lbrace
\bX^\T\beta^*\neq 0
\right\rbrace
+ \mathbbm{1}\left\lbrace
Y\bX^{\T}\beta < 0 
\right\rbrace \mathbbm{1}\left\lbrace
\bX^\T\beta^*= 0
\right\rbrace\right).
\end{equation*}
This completes the proof.

\subsection{Proof of Theorem \ref{thm:adaptive}}
Note that
	\begin{eqnarray*}
		P_n\left\lbrace 
		M_n(\beta_n^*) \notin \mathfrak{J}_{n, 1-\omega}
		\right\rbrace &=& 
		P_n\left\lbrace 
		M_n(\beta_n^*) \notin \mathfrak{J}_{n, 1-\omega},\,
		\beta_n^*\in \mathfrak{F}_{n,1-\eta}
		\right\rbrace +
		P_n\left\lbrace 
		M_n(\beta_n^*) \notin \mathfrak{J}_{n, 1-\omega},\,
		\beta_n^*\notin \mathfrak{F}_{n,1-\eta}
		\right\rbrace \\
		&\le& P_n\left\lbrace 
		M_n(\beta_n^*) \notin \mathfrak{W}_{n,1-\alpha}(\beta_n^*)\right\rbrace 
		+ P_n\left( \beta_n^*\notin \mathfrak{F}_{n,1-\eta}\right)\\
		&\le & \alpha + \eta + o(1)
	\end{eqnarray*}
where the last inequality follows from Lemma \ref{lemmyTheLemma}, the fact that $M_n(\beta_n^*) = \mathbb{G}_n(\beta^*_n,\beta^*_n)$, and $\sqrt n (\widehat{\beta}_n-\beta_n^*)\stackrel{d}{\to} N(0, \Sigma(\beta^*))$. The result follows immediately.
\end{document}